\documentclass{jfm}
\usepackage{graphicx}
\usepackage{newtxtext}
\usepackage{newtxmath}
\usepackage{natbib}
\usepackage[colorlinks=true, urlcolor=blue, linkcolor=blue,citecolor=blue]{hyperref}

\newcommand{\RomanNumeralCaps}[1]
\linenumbers
\usepackage{color}
\usepackage{xcolor}
\definecolor{AliceBlue}{rgb}{0.94,0.97,1.00}
\definecolor{AntiqueWhite1}{rgb}{1.00,0.94,0.86}
\definecolor{AntiqueWhite2}{rgb}{0.93,0.87,0.80}
\definecolor{AntiqueWhite3}{rgb}{0.80,0.75,0.69}
\definecolor{AntiqueWhite4}{rgb}{0.55,0.51,0.47}
\definecolor{AntiqueWhite}{rgb}{0.98,0.92,0.84}
\definecolor{BlanchedAlmond}{rgb}{1.00,0.92,0.80}
\definecolor{BlueViolet}{rgb}{0.54,0.17,0.89}
\definecolor{CadetBlue1}{rgb}{0.60,0.96,1.00}
\definecolor{CadetBlue2}{rgb}{0.56,0.90,0.93}
\definecolor{CadetBlue3}{rgb}{0.48,0.77,0.80}
\definecolor{CadetBlue4}{rgb}{0.33,0.53,0.55}
\definecolor{CadetBlue}{rgb}{0.37,0.62,0.63}
\definecolor{CornflowerBlue}{rgb}{0.39,0.58,0.93}
\definecolor{DarkBlue}{rgb}{0.00,0.00,0.55}
\definecolor{DarkCyan}{rgb}{0.00,0.55,0.55}
\definecolor{DarkGoldenrod1}{rgb}{1.00,0.73,0.06}
\definecolor{DarkGoldenrod2}{rgb}{0.93,0.68,0.05}
\definecolor{DarkGoldenrod3}{rgb}{0.80,0.58,0.05}
\definecolor{DarkGoldenrod4}{rgb}{0.55,0.40,0.03}
\definecolor{DarkGoldenrod}{rgb}{0.72,0.53,0.04}
\definecolor{DarkGray}{rgb}{0.66,0.66,0.66}
\definecolor{DarkGreen}{rgb}{0.00,0.39,0.00}
\definecolor{DarkGrey}{rgb}{0.66,0.66,0.66}
\definecolor{DarkKhaki}{rgb}{0.74,0.72,0.42}
\definecolor{DarkMagenta}{rgb}{0.55,0.00,0.55}
\definecolor{DarkOliveGreen1}{rgb}{0.79,1.00,0.44}
\definecolor{DarkOliveGreen2}{rgb}{0.74,0.93,0.41}
\definecolor{DarkOliveGreen3}{rgb}{0.64,0.80,0.35}
\definecolor{DarkOliveGreen4}{rgb}{0.43,0.55,0.24}
\definecolor{DarkOliveGreen}{rgb}{0.33,0.42,0.18}
\definecolor{DarkOrange1}{rgb}{1.00,0.50,0.00}
\definecolor{DarkOrange2}{rgb}{0.93,0.46,0.00}
\definecolor{DarkOrange3}{rgb}{0.80,0.40,0.00}
\definecolor{DarkOrange4}{rgb}{0.55,0.27,0.00}
\definecolor{DarkOrange}{rgb}{1.00,0.55,0.00}
\definecolor{DarkOrchid1}{rgb}{0.75,0.24,1.00}
\definecolor{DarkOrchid2}{rgb}{0.70,0.23,0.93}
\definecolor{DarkOrchid3}{rgb}{0.60,0.20,0.80}
\definecolor{DarkOrchid4}{rgb}{0.41,0.13,0.55}
\definecolor{DarkOrchid}{rgb}{0.60,0.20,0.80}
\definecolor{DarkRed}{rgb}{0.55,0.00,0.00}
\definecolor{DarkSalmon}{rgb}{0.91,0.59,0.48}
\definecolor{DarkSeaGreen1}{rgb}{0.76,1.00,0.76}
\definecolor{DarkSeaGreen2}{rgb}{0.71,0.93,0.71}
\definecolor{DarkSeaGreen3}{rgb}{0.61,0.80,0.61}
\definecolor{DarkSeaGreen4}{rgb}{0.41,0.55,0.41}
\definecolor{DarkSeaGreen}{rgb}{0.56,0.74,0.56}
\definecolor{DarkSlateBlue}{rgb}{0.28,0.24,0.55}
\definecolor{DarkSlateGray1}{rgb}{0.59,1.00,1.00}
\definecolor{DarkSlateGray2}{rgb}{0.55,0.93,0.93}
\definecolor{DarkSlateGray3}{rgb}{0.47,0.80,0.80}
\definecolor{DarkSlateGray4}{rgb}{0.32,0.55,0.55}
\definecolor{DarkSlateGray}{rgb}{0.18,0.31,0.31}
\definecolor{DarkSlateGrey}{rgb}{0.18,0.31,0.31}
\definecolor{DarkTurquoise}{rgb}{0.00,0.81,0.82}
\definecolor{DarkViolet}{rgb}{0.58,0.00,0.83}
\definecolor{DeepPink1}{rgb}{1.00,0.08,0.58}
\definecolor{DeepPink2}{rgb}{0.93,0.07,0.54}
\definecolor{DeepPink3}{rgb}{0.80,0.06,0.46}
\definecolor{DeepPink4}{rgb}{0.55,0.04,0.31}
\definecolor{DeepPink}{rgb}{1.00,0.08,0.58}
\definecolor{DeepSkyBlue1}{rgb}{0.00,0.75,1.00}
\definecolor{DeepSkyBlue2}{rgb}{0.00,0.70,0.93}
\definecolor{DeepSkyBlue3}{rgb}{0.00,0.60,0.80}
\definecolor{DeepSkyBlue4}{rgb}{0.00,0.41,0.55}
\definecolor{DeepSkyBlue}{rgb}{0.00,0.75,1.00}
\definecolor{DimGray}{rgb}{0.41,0.41,0.41}
\definecolor{DimGrey}{rgb}{0.41,0.41,0.41}
\definecolor{DodgerBlue1}{rgb}{0.12,0.56,1.00}
\definecolor{DodgerBlue2}{rgb}{0.11,0.53,0.93}
\definecolor{DodgerBlue3}{rgb}{0.09,0.45,0.80}
\definecolor{DodgerBlue4}{rgb}{0.06,0.31,0.55}
\definecolor{DodgerBlue}{rgb}{0.12,0.56,1.00}
\definecolor{FloralWhite}{rgb}{1.00,0.98,0.94}
\definecolor{ForestGreen}{rgb}{0.13,0.55,0.13}
\definecolor{GhostWhite}{rgb}{0.97,0.97,1.00}
\definecolor{GreenYellow}{rgb}{0.68,1.00,0.18}
\definecolor{HotPink1}{rgb}{1.00,0.43,0.71}
\definecolor{HotPink2}{rgb}{0.93,0.42,0.65}
\definecolor{HotPink3}{rgb}{0.80,0.38,0.56}
\definecolor{HotPink4}{rgb}{0.55,0.23,0.38}
\definecolor{HotPink}{rgb}{1.00,0.41,0.71}
\definecolor{IndianRed1}{rgb}{1.00,0.42,0.42}
\definecolor{IndianRed2}{rgb}{0.93,0.39,0.39}
\definecolor{IndianRed3}{rgb}{0.80,0.33,0.33}
\definecolor{IndianRed4}{rgb}{0.55,0.23,0.23}
\definecolor{IndianRed}{rgb}{0.80,0.36,0.36}
\definecolor{LavenderBlush1}{rgb}{1.00,0.94,0.96}
\definecolor{LavenderBlush2}{rgb}{0.93,0.88,0.90}
\definecolor{LavenderBlush3}{rgb}{0.80,0.76,0.77}
\definecolor{LavenderBlush4}{rgb}{0.55,0.51,0.53}
\definecolor{LavenderBlush}{rgb}{1.00,0.94,0.96}
\definecolor{LawnGreen}{rgb}{0.49,0.99,0.00}
\definecolor{LemonChiffon1}{rgb}{1.00,0.98,0.80}
\definecolor{LemonChiffon2}{rgb}{0.93,0.91,0.75}
\definecolor{LemonChiffon3}{rgb}{0.80,0.79,0.65}
\definecolor{LemonChiffon4}{rgb}{0.55,0.54,0.44}
\definecolor{LemonChiffon}{rgb}{1.00,0.98,0.80}
\definecolor{LightBlue1}{rgb}{0.75,0.94,1.00}
\definecolor{LightBlue2}{rgb}{0.70,0.87,0.93}
\definecolor{LightBlue3}{rgb}{0.60,0.75,0.80}
\definecolor{LightBlue4}{rgb}{0.41,0.51,0.55}
\definecolor{LightBlue}{rgb}{0.68,0.85,0.90}
\definecolor{LightCoral}{rgb}{0.94,0.50,0.50}
\definecolor{LightCyan1}{rgb}{0.88,1.00,1.00}
\definecolor{LightCyan2}{rgb}{0.82,0.93,0.93}
\definecolor{LightCyan3}{rgb}{0.71,0.80,0.80}
\definecolor{LightCyan4}{rgb}{0.48,0.55,0.55}
\definecolor{LightCyan}{rgb}{0.88,1.00,1.00}
\definecolor{LightGoldenrod1}{rgb}{1.00,0.93,0.55}
\definecolor{LightGoldenrod2}{rgb}{0.93,0.86,0.51}
\definecolor{LightGoldenrod3}{rgb}{0.80,0.75,0.44}
\definecolor{LightGoldenrod4}{rgb}{0.55,0.51,0.30}
\definecolor{LightGoldenrodYellow}{rgb}{0.98,0.98,0.82}
\definecolor{LightGoldenrod}{rgb}{0.93,0.87,0.51}
\definecolor{LightGray}{rgb}{0.83,0.83,0.83}
\definecolor{LightGreen}{rgb}{0.56,0.93,0.56}
\definecolor{LightGrey}{rgb}{0.83,0.83,0.83}
\definecolor{LightPink1}{rgb}{1.00,0.68,0.73}
\definecolor{LightPink2}{rgb}{0.93,0.64,0.68}
\definecolor{LightPink3}{rgb}{0.80,0.55,0.58}
\definecolor{LightPink4}{rgb}{0.55,0.37,0.40}
\definecolor{LightPink}{rgb}{1.00,0.71,0.76}
\definecolor{LightSalmon1}{rgb}{1.00,0.63,0.48}
\definecolor{LightSalmon2}{rgb}{0.93,0.58,0.45}
\definecolor{LightSalmon3}{rgb}{0.80,0.51,0.38}
\definecolor{LightSalmon4}{rgb}{0.55,0.34,0.26}
\definecolor{LightSalmon}{rgb}{1.00,0.63,0.48}
\definecolor{LightSeaGreen}{rgb}{0.13,0.70,0.67}
\definecolor{LightSkyBlue1}{rgb}{0.69,0.89,1.00}
\definecolor{LightSkyBlue2}{rgb}{0.64,0.83,0.93}
\definecolor{LightSkyBlue3}{rgb}{0.55,0.71,0.80}
\definecolor{LightSkyBlue4}{rgb}{0.38,0.48,0.55}
\definecolor{LightSkyBlue}{rgb}{0.53,0.81,0.98}
\definecolor{LightSlateBlue}{rgb}{0.52,0.44,1.00}
\definecolor{LightSlateGray}{rgb}{0.47,0.53,0.60}
\definecolor{LightSlateGrey}{rgb}{0.47,0.53,0.60}
\definecolor{LightSteelBlue1}{rgb}{0.79,0.88,1.00}
\definecolor{LightSteelBlue2}{rgb}{0.74,0.82,0.93}
\definecolor{LightSteelBlue3}{rgb}{0.64,0.71,0.80}
\definecolor{LightSteelBlue4}{rgb}{0.43,0.48,0.55}
\definecolor{LightSteelBlue}{rgb}{0.69,0.77,0.87}
\definecolor{LightYellow1}{rgb}{1.00,1.00,0.88}
\definecolor{LightYellow2}{rgb}{0.93,0.93,0.82}
\definecolor{LightYellow3}{rgb}{0.80,0.80,0.71}
\definecolor{LightYellow4}{rgb}{0.55,0.55,0.48}
\definecolor{LightYellow}{rgb}{1.00,1.00,0.88}
\definecolor{LimeGreen}{rgb}{0.20,0.80,0.20}
\definecolor{MediumAquamarine}{rgb}{0.40,0.80,0.67}
\definecolor{MediumBlue}{rgb}{0.00,0.00,0.80}
\definecolor{MediumOrchid1}{rgb}{0.88,0.40,1.00}
\definecolor{MediumOrchid2}{rgb}{0.82,0.37,0.93}
\definecolor{MediumOrchid3}{rgb}{0.71,0.32,0.80}
\definecolor{MediumOrchid4}{rgb}{0.48,0.22,0.55}
\definecolor{MediumOrchid}{rgb}{0.73,0.33,0.83}
\definecolor{MediumPurple1}{rgb}{0.67,0.51,1.00}
\definecolor{MediumPurple2}{rgb}{0.62,0.47,0.93}
\definecolor{MediumPurple3}{rgb}{0.54,0.41,0.80}
\definecolor{MediumPurple4}{rgb}{0.36,0.28,0.55}
\definecolor{MediumPurple}{rgb}{0.58,0.44,0.86}
\definecolor{MediumSeaGreen}{rgb}{0.24,0.70,0.44}
\definecolor{MediumSlateBlue}{rgb}{0.48,0.41,0.93}
\definecolor{MediumSpringGreen}{rgb}{0.00,0.98,0.60}
\definecolor{MediumTurquoise}{rgb}{0.28,0.82,0.80}
\definecolor{MediumVioletRed}{rgb}{0.78,0.08,0.52}
\definecolor{MidnightBlue}{rgb}{0.10,0.10,0.44}
\definecolor{MintCream}{rgb}{0.96,1.00,0.98}
\definecolor{MistyRose1}{rgb}{1.00,0.89,0.88}
\definecolor{MistyRose2}{rgb}{0.93,0.84,0.82}
\definecolor{MistyRose3}{rgb}{0.80,0.72,0.71}
\definecolor{MistyRose4}{rgb}{0.55,0.49,0.48}
\definecolor{MistyRose}{rgb}{1.00,0.89,0.88}
\definecolor{NavajoWhite1}{rgb}{1.00,0.87,0.68}
\definecolor{NavajoWhite2}{rgb}{0.93,0.81,0.63}
\definecolor{NavajoWhite3}{rgb}{0.80,0.70,0.55}
\definecolor{NavajoWhite4}{rgb}{0.55,0.47,0.37}
\definecolor{NavajoWhite}{rgb}{1.00,0.87,0.68}
\definecolor{NavyBlue}{rgb}{0.00,0.00,0.50}
\definecolor{OldLace}{rgb}{0.99,0.96,0.90}
\definecolor{OliveDrab1}{rgb}{0.75,1.00,0.24}
\definecolor{OliveDrab2}{rgb}{0.70,0.93,0.23}
\definecolor{OliveDrab3}{rgb}{0.60,0.80,0.20}
\definecolor{OliveDrab4}{rgb}{0.41,0.55,0.13}
\definecolor{OliveDrab}{rgb}{0.42,0.56,0.14}
\definecolor{OrangeRed1}{rgb}{1.00,0.27,0.00}
\definecolor{OrangeRed2}{rgb}{0.93,0.25,0.00}
\definecolor{OrangeRed3}{rgb}{0.80,0.22,0.00}
\definecolor{OrangeRed4}{rgb}{0.55,0.15,0.00}
\definecolor{OrangeRed}{rgb}{1.00,0.27,0.00}
\definecolor{PaleGoldenrod}{rgb}{0.93,0.91,0.67}
\definecolor{PaleGreen1}{rgb}{0.60,1.00,0.60}
\definecolor{PaleGreen2}{rgb}{0.56,0.93,0.56}
\definecolor{PaleGreen3}{rgb}{0.49,0.80,0.49}
\definecolor{PaleGreen4}{rgb}{0.33,0.55,0.33}
\definecolor{PaleGreen}{rgb}{0.60,0.98,0.60}
\definecolor{PaleTurquoise1}{rgb}{0.73,1.00,1.00}
\definecolor{PaleTurquoise2}{rgb}{0.68,0.93,0.93}
\definecolor{PaleTurquoise3}{rgb}{0.59,0.80,0.80}
\definecolor{PaleTurquoise4}{rgb}{0.40,0.55,0.55}
\definecolor{PaleTurquoise}{rgb}{0.69,0.93,0.93}
\definecolor{PaleVioletRed1}{rgb}{1.00,0.51,0.67}
\definecolor{PaleVioletRed2}{rgb}{0.93,0.47,0.62}
\definecolor{PaleVioletRed3}{rgb}{0.80,0.41,0.54}
\definecolor{PaleVioletRed4}{rgb}{0.55,0.28,0.36}
\definecolor{PaleVioletRed}{rgb}{0.86,0.44,0.58}
\definecolor{PapayaWhip}{rgb}{1.00,0.94,0.84}
\definecolor{PeachPuff1}{rgb}{1.00,0.85,0.73}
\definecolor{PeachPuff2}{rgb}{0.93,0.80,0.68}
\definecolor{PeachPuff3}{rgb}{0.80,0.69,0.58}
\definecolor{PeachPuff4}{rgb}{0.55,0.47,0.40}
\definecolor{PeachPuff}{rgb}{1.00,0.85,0.73}
\definecolor{PowderBlue}{rgb}{0.69,0.88,0.90}
\definecolor{RosyBrown1}{rgb}{1.00,0.76,0.76}
\definecolor{RosyBrown2}{rgb}{0.93,0.71,0.71}
\definecolor{RosyBrown3}{rgb}{0.80,0.61,0.61}
\definecolor{RosyBrown4}{rgb}{0.55,0.41,0.41}
\definecolor{RosyBrown}{rgb}{0.74,0.56,0.56}
\definecolor{RoyalBlue1}{rgb}{0.28,0.46,1.00}
\definecolor{RoyalBlue2}{rgb}{0.26,0.43,0.93}
\definecolor{RoyalBlue3}{rgb}{0.23,0.37,0.80}
\definecolor{RoyalBlue4}{rgb}{0.15,0.25,0.55}
\definecolor{RoyalBlue}{rgb}{0.25,0.41,0.88}
\definecolor{SaddleBrown}{rgb}{0.55,0.27,0.07}
\definecolor{SandyBrown}{rgb}{0.96,0.64,0.38}
\definecolor{SeaGreen1}{rgb}{0.33,1.00,0.62}
\definecolor{SeaGreen2}{rgb}{0.31,0.93,0.58}
\definecolor{SeaGreen3}{rgb}{0.26,0.80,0.50}
\definecolor{SeaGreen4}{rgb}{0.18,0.55,0.34}
\definecolor{SeaGreen}{rgb}{0.18,0.55,0.34}
\definecolor{SkyBlue1}{rgb}{0.53,0.81,1.00}
\definecolor{SkyBlue2}{rgb}{0.49,0.75,0.93}
\definecolor{SkyBlue3}{rgb}{0.42,0.65,0.80}
\definecolor{SkyBlue4}{rgb}{0.29,0.44,0.55}
\definecolor{SkyBlue}{rgb}{0.53,0.81,0.92}
\definecolor{SlateBlue1}{rgb}{0.51,0.44,1.00}
\definecolor{SlateBlue2}{rgb}{0.48,0.40,0.93}
\definecolor{SlateBlue3}{rgb}{0.41,0.35,0.80}
\definecolor{SlateBlue4}{rgb}{0.28,0.24,0.55}
\definecolor{SlateBlue}{rgb}{0.42,0.35,0.80}
\definecolor{SlateGray1}{rgb}{0.78,0.89,1.00}
\definecolor{SlateGray2}{rgb}{0.73,0.83,0.93}
\definecolor{SlateGray3}{rgb}{0.62,0.71,0.80}
\definecolor{SlateGray4}{rgb}{0.42,0.48,0.55}
\definecolor{SlateGray}{rgb}{0.44,0.50,0.56}
\definecolor{SlateGrey}{rgb}{0.44,0.50,0.56}
\definecolor{SpringGreen1}{rgb}{0.00,1.00,0.50}
\definecolor{SpringGreen2}{rgb}{0.00,0.93,0.46}
\definecolor{SpringGreen3}{rgb}{0.00,0.80,0.40}
\definecolor{SpringGreen4}{rgb}{0.00,0.55,0.27}
\definecolor{SpringGreen}{rgb}{0.00,1.00,0.50}
\definecolor{SteelBlue1}{rgb}{0.39,0.72,1.00}
\definecolor{SteelBlue2}{rgb}{0.36,0.67,0.93}
\definecolor{SteelBlue3}{rgb}{0.31,0.58,0.80}
\definecolor{SteelBlue4}{rgb}{0.21,0.39,0.55}
\definecolor{SteelBlue}{rgb}{0.27,0.51,0.71}
\definecolor{VioletRed1}{rgb}{1.00,0.24,0.59}
\definecolor{VioletRed2}{rgb}{0.93,0.23,0.55}
\definecolor{VioletRed3}{rgb}{0.80,0.20,0.47}
\definecolor{VioletRed4}{rgb}{0.55,0.13,0.32}
\definecolor{VioletRed}{rgb}{0.82,0.13,0.56}
\definecolor{WhiteSmoke}{rgb}{0.96,0.96,0.96}
\definecolor{YellowGreen}{rgb}{0.60,0.80,0.20}
\definecolor{aliceblue}{rgb}{0.94,0.97,1.00}
\definecolor{antiquewhite}{rgb}{0.98,0.92,0.84}
\definecolor{aquamarine1}{rgb}{0.50,1.00,0.83}
\definecolor{aquamarine2}{rgb}{0.46,0.93,0.78}
\definecolor{aquamarine3}{rgb}{0.40,0.80,0.67}
\definecolor{aquamarine4}{rgb}{0.27,0.55,0.45}
\definecolor{aquamarine}{rgb}{0.50,1.00,0.83}
\definecolor{azure1}{rgb}{0.94,1.00,1.00}
\definecolor{azure2}{rgb}{0.88,0.93,0.93}
\definecolor{azure3}{rgb}{0.76,0.80,0.80}
\definecolor{azure4}{rgb}{0.51,0.55,0.55}
\definecolor{azure}{rgb}{0.94,1.00,1.00}
\definecolor{beige}{rgb}{0.96,0.96,0.86}
\definecolor{bisque1}{rgb}{1.00,0.89,0.77}
\definecolor{bisque2}{rgb}{0.93,0.84,0.72}
\definecolor{bisque3}{rgb}{0.80,0.72,0.62}
\definecolor{bisque4}{rgb}{0.55,0.49,0.42}
\definecolor{bisque}{rgb}{1.00,0.89,0.77}
\definecolor{black}{rgb}{0.00,0.00,0.00}
\definecolor{blanchedalmond}{rgb}{1.00,0.92,0.80}
\definecolor{blue1}{rgb}{0.00,0.00,1.00}
\definecolor{blue2}{rgb}{0.00,0.00,0.93}
\definecolor{blue3}{rgb}{0.00,0.00,0.80}
\definecolor{blue4}{rgb}{0.00,0.00,0.55}
\definecolor{blueviolet}{rgb}{0.54,0.17,0.89}
\definecolor{blue}{rgb}{0.00,0.00,1.00}
\definecolor{brown1}{rgb}{1.00,0.25,0.25}
\definecolor{brown2}{rgb}{0.93,0.23,0.23}
\definecolor{brown3}{rgb}{0.80,0.20,0.20}
\definecolor{brown4}{rgb}{0.55,0.14,0.14}
\definecolor{brown}{rgb}{0.65,0.16,0.16}
\definecolor{burlywood1}{rgb}{1.00,0.83,0.61}
\definecolor{burlywood2}{rgb}{0.93,0.77,0.57}
\definecolor{burlywood3}{rgb}{0.80,0.67,0.49}
\definecolor{burlywood4}{rgb}{0.55,0.45,0.33}
\definecolor{burlywood}{rgb}{0.87,0.72,0.53}
\definecolor{cadetblue}{rgb}{0.37,0.62,0.63}
\definecolor{chartreuse1}{rgb}{0.50,1.00,0.00}
\definecolor{chartreuse2}{rgb}{0.46,0.93,0.00}
\definecolor{chartreuse3}{rgb}{0.40,0.80,0.00}
\definecolor{chartreuse4}{rgb}{0.27,0.55,0.00}
\definecolor{chartreuse}{rgb}{0.50,1.00,0.00}
\definecolor{chocolate1}{rgb}{1.00,0.50,0.14}
\definecolor{chocolate2}{rgb}{0.93,0.46,0.13}
\definecolor{chocolate3}{rgb}{0.80,0.40,0.11}
\definecolor{chocolate4}{rgb}{0.55,0.27,0.07}
\definecolor{chocolate}{rgb}{0.82,0.41,0.12}
\definecolor{coral1}{rgb}{1.00,0.45,0.34}
\definecolor{coral2}{rgb}{0.93,0.42,0.31}
\definecolor{coral3}{rgb}{0.80,0.36,0.27}
\definecolor{coral4}{rgb}{0.55,0.24,0.18}
\definecolor{coral}{rgb}{1.00,0.50,0.31}
\definecolor{cornflowerblue}{rgb}{0.39,0.58,0.93}
\definecolor{cornsilk1}{rgb}{1.00,0.97,0.86}
\definecolor{cornsilk2}{rgb}{0.93,0.91,0.80}
\definecolor{cornsilk3}{rgb}{0.80,0.78,0.69}
\definecolor{cornsilk4}{rgb}{0.55,0.53,0.47}
\definecolor{cornsilk}{rgb}{1.00,0.97,0.86}
\definecolor{cyan1}{rgb}{0.00,1.00,1.00}
\definecolor{cyan2}{rgb}{0.00,0.93,0.93}
\definecolor{cyan3}{rgb}{0.00,0.80,0.80}
\definecolor{cyan4}{rgb}{0.00,0.55,0.55}
\definecolor{cyan}{rgb}{0.00,1.00,1.00}
\definecolor{darkblue}{rgb}{0.00,0.00,0.55}
\definecolor{darkcyan}{rgb}{0.00,0.55,0.55}
\definecolor{darkgoldenrod}{rgb}{0.72,0.53,0.04}
\definecolor{darkgray}{rgb}{0.66,0.66,0.66}
\definecolor{darkgreen}{rgb}{0.00,0.39,0.00}
\definecolor{darkgrey}{rgb}{0.66,0.66,0.66}
\definecolor{darkkhaki}{rgb}{0.74,0.72,0.42}
\definecolor{darkmagenta}{rgb}{0.55,0.00,0.55}
\definecolor{darkolive}{rgb}{0.33,0.42,0.18}
\definecolor{darkorange}{rgb}{1.00,0.55,0.00}
\definecolor{darkorchid}{rgb}{0.60,0.20,0.80}
\definecolor{darkred}{rgb}{0.55,0.00,0.00}
\definecolor{darksalmon}{rgb}{0.91,0.59,0.48}
\definecolor{darksea}{rgb}{0.56,0.74,0.56}
\definecolor{darkslate}{rgb}{0.18,0.31,0.31}
\definecolor{darkslate}{rgb}{0.18,0.31,0.31}
\definecolor{darkslate}{rgb}{0.28,0.24,0.55}
\definecolor{darkturquoise}{rgb}{0.00,0.81,0.82}
\definecolor{darkviolet}{rgb}{0.58,0.00,0.83}
\definecolor{deeppink}{rgb}{1.00,0.08,0.58}
\definecolor{deepsky}{rgb}{0.00,0.75,1.00}
\definecolor{dimgray}{rgb}{0.41,0.41,0.41}
\definecolor{dimgrey}{rgb}{0.41,0.41,0.41}
\definecolor{dodgerblue}{rgb}{0.12,0.56,1.00}
\definecolor{firebrick1}{rgb}{1.00,0.19,0.19}
\definecolor{firebrick2}{rgb}{0.93,0.17,0.17}
\definecolor{firebrick3}{rgb}{0.80,0.15,0.15}
\definecolor{firebrick4}{rgb}{0.55,0.10,0.10}
\definecolor{firebrick}{rgb}{0.70,0.13,0.13}
\definecolor{floralwhite}{rgb}{1.00,0.98,0.94}
\definecolor{forestgreen}{rgb}{0.13,0.55,0.13}
\definecolor{gainsboro}{rgb}{0.86,0.86,0.86}
\definecolor{ghostwhite}{rgb}{0.97,0.97,1.00}
\definecolor{gold1}{rgb}{1.00,0.84,0.00}
\definecolor{gold2}{rgb}{0.93,0.79,0.00}
\definecolor{gold3}{rgb}{0.80,0.68,0.00}
\definecolor{gold4}{rgb}{0.55,0.46,0.00}
\definecolor{goldenrod1}{rgb}{1.00,0.76,0.15}
\definecolor{goldenrod2}{rgb}{0.93,0.71,0.13}
\definecolor{goldenrod3}{rgb}{0.80,0.61,0.11}
\definecolor{goldenrod4}{rgb}{0.55,0.41,0.08}
\definecolor{goldenrod}{rgb}{0.85,0.65,0.13}
\definecolor{gold}{rgb}{1.00,0.84,0.00}
\definecolor{gray0}{rgb}{0.00,0.00,0.00}
\definecolor{gray100}{rgb}{1.00,1.00,1.00}
\definecolor{gray10}{rgb}{0.10,0.10,0.10}
\definecolor{gray11}{rgb}{0.11,0.11,0.11}
\definecolor{gray12}{rgb}{0.12,0.12,0.12}
\definecolor{gray13}{rgb}{0.13,0.13,0.13}
\definecolor{gray14}{rgb}{0.14,0.14,0.14}
\definecolor{gray15}{rgb}{0.15,0.15,0.15}
\definecolor{gray16}{rgb}{0.16,0.16,0.16}
\definecolor{gray17}{rgb}{0.17,0.17,0.17}
\definecolor{gray18}{rgb}{0.18,0.18,0.18}
\definecolor{gray19}{rgb}{0.19,0.19,0.19}
\definecolor{gray1}{rgb}{0.01,0.01,0.01}
\definecolor{gray20}{rgb}{0.20,0.20,0.20}
\definecolor{gray21}{rgb}{0.21,0.21,0.21}
\definecolor{gray22}{rgb}{0.22,0.22,0.22}
\definecolor{gray23}{rgb}{0.23,0.23,0.23}
\definecolor{gray24}{rgb}{0.24,0.24,0.24}
\definecolor{gray25}{rgb}{0.25,0.25,0.25}
\definecolor{gray26}{rgb}{0.26,0.26,0.26}
\definecolor{gray27}{rgb}{0.27,0.27,0.27}
\definecolor{gray28}{rgb}{0.28,0.28,0.28}
\definecolor{gray29}{rgb}{0.29,0.29,0.29}
\definecolor{gray2}{rgb}{0.02,0.02,0.02}
\definecolor{gray30}{rgb}{0.30,0.30,0.30}
\definecolor{gray31}{rgb}{0.31,0.31,0.31}
\definecolor{gray32}{rgb}{0.32,0.32,0.32}
\definecolor{gray33}{rgb}{0.33,0.33,0.33}
\definecolor{gray34}{rgb}{0.34,0.34,0.34}
\definecolor{gray35}{rgb}{0.35,0.35,0.35}
\definecolor{gray36}{rgb}{0.36,0.36,0.36}
\definecolor{gray37}{rgb}{0.37,0.37,0.37}
\definecolor{gray38}{rgb}{0.38,0.38,0.38}
\definecolor{gray39}{rgb}{0.39,0.39,0.39}
\definecolor{gray3}{rgb}{0.03,0.03,0.03}
\definecolor{gray40}{rgb}{0.40,0.40,0.40}
\definecolor{gray41}{rgb}{0.41,0.41,0.41}
\definecolor{gray42}{rgb}{0.42,0.42,0.42}
\definecolor{gray43}{rgb}{0.43,0.43,0.43}
\definecolor{gray44}{rgb}{0.44,0.44,0.44}
\definecolor{gray45}{rgb}{0.45,0.45,0.45}
\definecolor{gray46}{rgb}{0.46,0.46,0.46}
\definecolor{gray47}{rgb}{0.47,0.47,0.47}
\definecolor{gray48}{rgb}{0.48,0.48,0.48}
\definecolor{gray49}{rgb}{0.49,0.49,0.49}
\definecolor{gray4}{rgb}{0.04,0.04,0.04}
\definecolor{gray50}{rgb}{0.50,0.50,0.50}
\definecolor{gray51}{rgb}{0.51,0.51,0.51}
\definecolor{gray52}{rgb}{0.52,0.52,0.52}
\definecolor{gray53}{rgb}{0.53,0.53,0.53}
\definecolor{gray54}{rgb}{0.54,0.54,0.54}
\definecolor{gray55}{rgb}{0.55,0.55,0.55}
\definecolor{gray56}{rgb}{0.56,0.56,0.56}
\definecolor{gray57}{rgb}{0.57,0.57,0.57}
\definecolor{gray58}{rgb}{0.58,0.58,0.58}
\definecolor{gray59}{rgb}{0.59,0.59,0.59}
\definecolor{gray5}{rgb}{0.05,0.05,0.05}
\definecolor{gray60}{rgb}{0.60,0.60,0.60}
\definecolor{gray61}{rgb}{0.61,0.61,0.61}
\definecolor{gray62}{rgb}{0.62,0.62,0.62}
\definecolor{gray63}{rgb}{0.63,0.63,0.63}
\definecolor{gray64}{rgb}{0.64,0.64,0.64}
\definecolor{gray65}{rgb}{0.65,0.65,0.65}
\definecolor{gray66}{rgb}{0.66,0.66,0.66}
\definecolor{gray67}{rgb}{0.67,0.67,0.67}
\definecolor{gray68}{rgb}{0.68,0.68,0.68}
\definecolor{gray69}{rgb}{0.69,0.69,0.69}
\definecolor{gray6}{rgb}{0.06,0.06,0.06}
\definecolor{gray70}{rgb}{0.70,0.70,0.70}
\definecolor{gray71}{rgb}{0.71,0.71,0.71}
\definecolor{gray72}{rgb}{0.72,0.72,0.72}
\definecolor{gray73}{rgb}{0.73,0.73,0.73}
\definecolor{gray74}{rgb}{0.74,0.74,0.74}
\definecolor{gray75}{rgb}{0.75,0.75,0.75}
\definecolor{gray76}{rgb}{0.76,0.76,0.76}
\definecolor{gray77}{rgb}{0.77,0.77,0.77}
\definecolor{gray78}{rgb}{0.78,0.78,0.78}
\definecolor{gray79}{rgb}{0.79,0.79,0.79}
\definecolor{gray7}{rgb}{0.07,0.07,0.07}
\definecolor{gray80}{rgb}{0.80,0.80,0.80}
\definecolor{gray81}{rgb}{0.81,0.81,0.81}
\definecolor{gray82}{rgb}{0.82,0.82,0.82}
\definecolor{gray83}{rgb}{0.83,0.83,0.83}
\definecolor{gray84}{rgb}{0.84,0.84,0.84}
\definecolor{gray85}{rgb}{0.85,0.85,0.85}
\definecolor{gray86}{rgb}{0.86,0.86,0.86}
\definecolor{gray87}{rgb}{0.87,0.87,0.87}
\definecolor{gray88}{rgb}{0.88,0.88,0.88}
\definecolor{gray89}{rgb}{0.89,0.89,0.89}
\definecolor{gray8}{rgb}{0.08,0.08,0.08}
\definecolor{gray90}{rgb}{0.90,0.90,0.90}
\definecolor{gray91}{rgb}{0.91,0.91,0.91}
\definecolor{gray92}{rgb}{0.92,0.92,0.92}
\definecolor{gray93}{rgb}{0.93,0.93,0.93}
\definecolor{gray94}{rgb}{0.94,0.94,0.94}
\definecolor{gray95}{rgb}{0.95,0.95,0.95}
\definecolor{gray96}{rgb}{0.96,0.96,0.96}
\definecolor{gray97}{rgb}{0.97,0.97,0.97}
\definecolor{gray98}{rgb}{0.98,0.98,0.98}
\definecolor{gray99}{rgb}{0.99,0.99,0.99}
\definecolor{gray9}{rgb}{0.09,0.09,0.09}
\definecolor{gray}{rgb}{0.75,0.75,0.75}
\definecolor{green1}{rgb}{0.00,1.00,0.00}
\definecolor{green2}{rgb}{0.00,0.93,0.00}
\definecolor{green3}{rgb}{0.00,0.80,0.00}
\definecolor{green4}{rgb}{0.00,0.55,0.00}
\definecolor{greenyellow}{rgb}{0.68,1.00,0.18}
\definecolor{green}{rgb}{0.00,1.00,0.00}
\definecolor{grey0}{rgb}{0.00,0.00,0.00}
\definecolor{grey100}{rgb}{1.00,1.00,1.00}
\definecolor{grey10}{rgb}{0.10,0.10,0.10}
\definecolor{grey11}{rgb}{0.11,0.11,0.11}
\definecolor{grey12}{rgb}{0.12,0.12,0.12}
\definecolor{grey13}{rgb}{0.13,0.13,0.13}
\definecolor{grey14}{rgb}{0.14,0.14,0.14}
\definecolor{grey15}{rgb}{0.15,0.15,0.15}
\definecolor{grey16}{rgb}{0.16,0.16,0.16}
\definecolor{grey17}{rgb}{0.17,0.17,0.17}
\definecolor{grey18}{rgb}{0.18,0.18,0.18}
\definecolor{grey19}{rgb}{0.19,0.19,0.19}
\definecolor{grey1}{rgb}{0.01,0.01,0.01}
\definecolor{grey20}{rgb}{0.20,0.20,0.20}
\definecolor{grey21}{rgb}{0.21,0.21,0.21}
\definecolor{grey22}{rgb}{0.22,0.22,0.22}
\definecolor{grey23}{rgb}{0.23,0.23,0.23}
\definecolor{grey24}{rgb}{0.24,0.24,0.24}
\definecolor{grey25}{rgb}{0.25,0.25,0.25}
\definecolor{grey26}{rgb}{0.26,0.26,0.26}
\definecolor{grey27}{rgb}{0.27,0.27,0.27}
\definecolor{grey28}{rgb}{0.28,0.28,0.28}
\definecolor{grey29}{rgb}{0.29,0.29,0.29}
\definecolor{grey2}{rgb}{0.02,0.02,0.02}
\definecolor{grey30}{rgb}{0.30,0.30,0.30}
\definecolor{grey31}{rgb}{0.31,0.31,0.31}
\definecolor{grey32}{rgb}{0.32,0.32,0.32}
\definecolor{grey33}{rgb}{0.33,0.33,0.33}
\definecolor{grey34}{rgb}{0.34,0.34,0.34}
\definecolor{grey35}{rgb}{0.35,0.35,0.35}
\definecolor{grey36}{rgb}{0.36,0.36,0.36}
\definecolor{grey37}{rgb}{0.37,0.37,0.37}
\definecolor{grey38}{rgb}{0.38,0.38,0.38}
\definecolor{grey39}{rgb}{0.39,0.39,0.39}
\definecolor{grey3}{rgb}{0.03,0.03,0.03}
\definecolor{grey40}{rgb}{0.40,0.40,0.40}
\definecolor{grey41}{rgb}{0.41,0.41,0.41}
\definecolor{grey42}{rgb}{0.42,0.42,0.42}
\definecolor{grey43}{rgb}{0.43,0.43,0.43}
\definecolor{grey44}{rgb}{0.44,0.44,0.44}
\definecolor{grey45}{rgb}{0.45,0.45,0.45}
\definecolor{grey46}{rgb}{0.46,0.46,0.46}
\definecolor{grey47}{rgb}{0.47,0.47,0.47}
\definecolor{grey48}{rgb}{0.48,0.48,0.48}
\definecolor{grey49}{rgb}{0.49,0.49,0.49}
\definecolor{grey4}{rgb}{0.04,0.04,0.04}
\definecolor{grey50}{rgb}{0.50,0.50,0.50}
\definecolor{grey51}{rgb}{0.51,0.51,0.51}
\definecolor{grey52}{rgb}{0.52,0.52,0.52}
\definecolor{grey53}{rgb}{0.53,0.53,0.53}
\definecolor{grey54}{rgb}{0.54,0.54,0.54}
\definecolor{grey55}{rgb}{0.55,0.55,0.55}
\definecolor{grey56}{rgb}{0.56,0.56,0.56}
\definecolor{grey57}{rgb}{0.57,0.57,0.57}
\definecolor{grey58}{rgb}{0.58,0.58,0.58}
\definecolor{grey59}{rgb}{0.59,0.59,0.59}
\definecolor{grey5}{rgb}{0.05,0.05,0.05}
\definecolor{grey60}{rgb}{0.60,0.60,0.60}
\definecolor{grey61}{rgb}{0.61,0.61,0.61}
\definecolor{grey62}{rgb}{0.62,0.62,0.62}
\definecolor{grey63}{rgb}{0.63,0.63,0.63}
\definecolor{grey64}{rgb}{0.64,0.64,0.64}
\definecolor{grey65}{rgb}{0.65,0.65,0.65}
\definecolor{grey66}{rgb}{0.66,0.66,0.66}
\definecolor{grey67}{rgb}{0.67,0.67,0.67}
\definecolor{grey68}{rgb}{0.68,0.68,0.68}
\definecolor{grey69}{rgb}{0.69,0.69,0.69}
\definecolor{grey6}{rgb}{0.06,0.06,0.06}
\definecolor{grey70}{rgb}{0.70,0.70,0.70}
\definecolor{grey71}{rgb}{0.71,0.71,0.71}
\definecolor{grey72}{rgb}{0.72,0.72,0.72}
\definecolor{grey73}{rgb}{0.73,0.73,0.73}
\definecolor{grey74}{rgb}{0.74,0.74,0.74}
\definecolor{grey75}{rgb}{0.75,0.75,0.75}
\definecolor{grey76}{rgb}{0.76,0.76,0.76}
\definecolor{grey77}{rgb}{0.77,0.77,0.77}
\definecolor{grey78}{rgb}{0.78,0.78,0.78}
\definecolor{grey79}{rgb}{0.79,0.79,0.79}
\definecolor{grey7}{rgb}{0.07,0.07,0.07}
\definecolor{grey80}{rgb}{0.80,0.80,0.80}
\definecolor{grey81}{rgb}{0.81,0.81,0.81}
\definecolor{grey82}{rgb}{0.82,0.82,0.82}
\definecolor{grey83}{rgb}{0.83,0.83,0.83}
\definecolor{grey84}{rgb}{0.84,0.84,0.84}
\definecolor{grey85}{rgb}{0.85,0.85,0.85}
\definecolor{grey86}{rgb}{0.86,0.86,0.86}
\definecolor{grey87}{rgb}{0.87,0.87,0.87}
\definecolor{grey88}{rgb}{0.88,0.88,0.88}
\definecolor{grey89}{rgb}{0.89,0.89,0.89}
\definecolor{grey8}{rgb}{0.08,0.08,0.08}
\definecolor{grey90}{rgb}{0.90,0.90,0.90}
\definecolor{grey91}{rgb}{0.91,0.91,0.91}
\definecolor{grey92}{rgb}{0.92,0.92,0.92}
\definecolor{grey93}{rgb}{0.93,0.93,0.93}
\definecolor{grey94}{rgb}{0.94,0.94,0.94}
\definecolor{grey95}{rgb}{0.95,0.95,0.95}
\definecolor{grey96}{rgb}{0.96,0.96,0.96}
\definecolor{grey97}{rgb}{0.97,0.97,0.97}
\definecolor{grey98}{rgb}{0.98,0.98,0.98}
\definecolor{grey99}{rgb}{0.99,0.99,0.99}
\definecolor{grey9}{rgb}{0.09,0.09,0.09}
\definecolor{grey}{rgb}{0.75,0.75,0.75}
\definecolor{honeydew1}{rgb}{0.94,1.00,0.94}
\definecolor{honeydew2}{rgb}{0.88,0.93,0.88}
\definecolor{honeydew3}{rgb}{0.76,0.80,0.76}
\definecolor{honeydew4}{rgb}{0.51,0.55,0.51}
\definecolor{honeydew}{rgb}{0.94,1.00,0.94}
\definecolor{hotpink}{rgb}{1.00,0.41,0.71}
\definecolor{indianred}{rgb}{0.80,0.36,0.36}
\definecolor{ivory1}{rgb}{1.00,1.00,0.94}
\definecolor{ivory2}{rgb}{0.93,0.93,0.88}
\definecolor{ivory3}{rgb}{0.80,0.80,0.76}
\definecolor{ivory4}{rgb}{0.55,0.55,0.51}
\definecolor{ivory}{rgb}{1.00,1.00,0.94}
\definecolor{khaki1}{rgb}{1.00,0.96,0.56}
\definecolor{khaki2}{rgb}{0.93,0.90,0.52}
\definecolor{khaki3}{rgb}{0.80,0.78,0.45}
\definecolor{khaki4}{rgb}{0.55,0.53,0.31}
\definecolor{khaki}{rgb}{0.94,0.90,0.55}
\definecolor{lavenderblush}{rgb}{1.00,0.94,0.96}
\definecolor{lavender}{rgb}{0.90,0.90,0.98}
\definecolor{lawngreen}{rgb}{0.49,0.99,0.00}
\definecolor{lemonchiffon}{rgb}{1.00,0.98,0.80}
\definecolor{lightblue}{rgb}{0.68,0.85,0.90}
\definecolor{lightcoral}{rgb}{0.94,0.50,0.50}
\definecolor{lightcyan}{rgb}{0.88,1.00,1.00}
\definecolor{lightgoldenrod}{rgb}{0.93,0.87,0.51}
\definecolor{lightgoldenrod}{rgb}{0.98,0.98,0.82}
\definecolor{lightgray}{rgb}{0.83,0.83,0.83}
\definecolor{lightgreen}{rgb}{0.56,0.93,0.56}
\definecolor{lightgrey}{rgb}{0.83,0.83,0.83}
\definecolor{lightpink}{rgb}{1.00,0.71,0.76}
\definecolor{lightsalmon}{rgb}{1.00,0.63,0.48}
\definecolor{lightsea}{rgb}{0.13,0.70,0.67}
\definecolor{lightsky}{rgb}{0.53,0.81,0.98}
\definecolor{lightslate}{rgb}{0.47,0.53,0.60}
\definecolor{lightslate}{rgb}{0.47,0.53,0.60}
\definecolor{lightslate}{rgb}{0.52,0.44,1.00}
\definecolor{lightsteel}{rgb}{0.69,0.77,0.87}
\definecolor{lightyellow}{rgb}{1.00,1.00,0.88}
\definecolor{limegreen}{rgb}{0.20,0.80,0.20}
\definecolor{linen}{rgb}{0.98,0.94,0.90}
\definecolor{magenta1}{rgb}{1.00,0.00,1.00}
\definecolor{magenta2}{rgb}{0.93,0.00,0.93}
\definecolor{magenta3}{rgb}{0.80,0.00,0.80}
\definecolor{magenta4}{rgb}{0.55,0.00,0.55}
\definecolor{magenta}{rgb}{1.00,0.00,1.00}
\definecolor{maroon1}{rgb}{1.00,0.20,0.70}
\definecolor{maroon2}{rgb}{0.93,0.19,0.65}
\definecolor{maroon3}{rgb}{0.80,0.16,0.56}
\definecolor{maroon4}{rgb}{0.55,0.11,0.38}
\definecolor{maroon}{rgb}{0.69,0.19,0.38}
\definecolor{mediumaquamarine}{rgb}{0.40,0.80,0.67}
\definecolor{mediumblue}{rgb}{0.00,0.00,0.80}
\definecolor{mediumorchid}{rgb}{0.73,0.33,0.83}
\definecolor{mediumpurple}{rgb}{0.58,0.44,0.86}
\definecolor{mediumsea}{rgb}{0.24,0.70,0.44}
\definecolor{mediumslate}{rgb}{0.48,0.41,0.93}
\definecolor{mediumspring}{rgb}{0.00,0.98,0.60}
\definecolor{mediumturquoise}{rgb}{0.28,0.82,0.80}
\definecolor{mediumviolet}{rgb}{0.78,0.08,0.52}
\definecolor{midnightblue}{rgb}{0.10,0.10,0.44}
\definecolor{mintcream}{rgb}{0.96,1.00,0.98}
\definecolor{mistyrose}{rgb}{1.00,0.89,0.88}
\definecolor{moccasin}{rgb}{1.00,0.89,0.71}
\definecolor{navajowhite}{rgb}{1.00,0.87,0.68}
\definecolor{navyblue}{rgb}{0.00,0.00,0.50}
\definecolor{navy}{rgb}{0.00,0.00,0.50}
\definecolor{oldlace}{rgb}{0.99,0.96,0.90}
\definecolor{olivedrab}{rgb}{0.42,0.56,0.14}
\definecolor{orange1}{rgb}{1.00,0.65,0.00}
\definecolor{orange2}{rgb}{0.93,0.60,0.00}
\definecolor{orange3}{rgb}{0.80,0.52,0.00}
\definecolor{orange4}{rgb}{0.55,0.35,0.00}
\definecolor{orangered}{rgb}{1.00,0.27,0.00}
\definecolor{orange}{rgb}{1.00,0.65,0.00}
\definecolor{orchid1}{rgb}{1.00,0.51,0.98}
\definecolor{orchid2}{rgb}{0.93,0.48,0.91}
\definecolor{orchid3}{rgb}{0.80,0.41,0.79}
\definecolor{orchid4}{rgb}{0.55,0.28,0.54}
\definecolor{orchid}{rgb}{0.85,0.44,0.84}
\definecolor{palegoldenrod}{rgb}{0.93,0.91,0.67}
\definecolor{palegreen}{rgb}{0.60,0.98,0.60}
\definecolor{paleturquoise}{rgb}{0.69,0.93,0.93}
\definecolor{paleviolet}{rgb}{0.86,0.44,0.58}
\definecolor{papayawhip}{rgb}{1.00,0.94,0.84}
\definecolor{peachpuff}{rgb}{1.00,0.85,0.73}
\definecolor{peru}{rgb}{0.80,0.52,0.25}
\definecolor{pink1}{rgb}{1.00,0.71,0.77}
\definecolor{pink2}{rgb}{0.93,0.66,0.72}
\definecolor{pink3}{rgb}{0.80,0.57,0.62}
\definecolor{pink4}{rgb}{0.55,0.39,0.42}
\definecolor{pink}{rgb}{1.00,0.75,0.80}
\definecolor{plum1}{rgb}{1.00,0.73,1.00}
\definecolor{plum2}{rgb}{0.93,0.68,0.93}
\definecolor{plum3}{rgb}{0.80,0.59,0.80}
\definecolor{plum4}{rgb}{0.55,0.40,0.55}
\definecolor{plum}{rgb}{0.87,0.63,0.87}
\definecolor{powderblue}{rgb}{0.69,0.88,0.90}
\definecolor{purple1}{rgb}{0.61,0.19,1.00}
\definecolor{purple2}{rgb}{0.57,0.17,0.93}
\definecolor{purple3}{rgb}{0.49,0.15,0.80}
\definecolor{purple4}{rgb}{0.33,0.10,0.55}
\definecolor{purple}{rgb}{0.63,0.13,0.94}
\definecolor{red1}{rgb}{1.00,0.00,0.00}
\definecolor{red2}{rgb}{0.93,0.00,0.00}
\definecolor{red3}{rgb}{0.80,0.00,0.00}
\definecolor{red4}{rgb}{0.55,0.00,0.00}
\definecolor{red}{rgb}{1.00,0.00,0.00}
\definecolor{rosybrown}{rgb}{0.74,0.56,0.56}
\definecolor{royalblue}{rgb}{0.25,0.41,0.88}
\definecolor{saddlebrown}{rgb}{0.55,0.27,0.07}
\definecolor{salmon1}{rgb}{1.00,0.55,0.41}
\definecolor{salmon2}{rgb}{0.93,0.51,0.38}
\definecolor{salmon3}{rgb}{0.80,0.44,0.33}
\definecolor{salmon4}{rgb}{0.55,0.30,0.22}
\definecolor{salmon}{rgb}{0.98,0.50,0.45}
\definecolor{sandybrown}{rgb}{0.96,0.64,0.38}
\definecolor{seagreen}{rgb}{0.18,0.55,0.34}
\definecolor{seashell1}{rgb}{1.00,0.96,0.93}
\definecolor{seashell2}{rgb}{0.93,0.90,0.87}
\definecolor{seashell3}{rgb}{0.80,0.77,0.75}
\definecolor{seashell4}{rgb}{0.55,0.53,0.51}
\definecolor{seashell}{rgb}{1.00,0.96,0.93}
\definecolor{sienna1}{rgb}{1.00,0.51,0.28}
\definecolor{sienna2}{rgb}{0.93,0.47,0.26}
\definecolor{sienna3}{rgb}{0.80,0.41,0.22}
\definecolor{sienna4}{rgb}{0.55,0.28,0.15}
\definecolor{sienna}{rgb}{0.63,0.32,0.18}
\definecolor{skyblue}{rgb}{0.53,0.81,0.92}
\definecolor{slateblue}{rgb}{0.42,0.35,0.80}
\definecolor{slategray}{rgb}{0.44,0.50,0.56}
\definecolor{slategrey}{rgb}{0.44,0.50,0.56}
\definecolor{snow1}{rgb}{1.00,0.98,0.98}
\definecolor{snow2}{rgb}{0.93,0.91,0.91}
\definecolor{snow3}{rgb}{0.80,0.79,0.79}
\definecolor{snow4}{rgb}{0.55,0.54,0.54}
\definecolor{snow}{rgb}{1.00,0.98,0.98}
\definecolor{springgreen}{rgb}{0.00,1.00,0.50}
\definecolor{steelblue}{rgb}{0.27,0.51,0.71}
\definecolor{tan1}{rgb}{1.00,0.65,0.31}
\definecolor{tan2}{rgb}{0.93,0.60,0.29}
\definecolor{tan3}{rgb}{0.80,0.52,0.25}
\definecolor{tan4}{rgb}{0.55,0.35,0.17}
\definecolor{tan}{rgb}{0.82,0.71,0.55}
\definecolor{thistle1}{rgb}{1.00,0.88,1.00}
\definecolor{thistle2}{rgb}{0.93,0.82,0.93}
\definecolor{thistle3}{rgb}{0.80,0.71,0.80}
\definecolor{thistle4}{rgb}{0.55,0.48,0.55}
\definecolor{thistle}{rgb}{0.85,0.75,0.85}
\definecolor{tomato1}{rgb}{1.00,0.39,0.28}
\definecolor{tomato2}{rgb}{0.93,0.36,0.26}
\definecolor{tomato3}{rgb}{0.80,0.31,0.22}
\definecolor{tomato4}{rgb}{0.55,0.21,0.15}
\definecolor{tomato}{rgb}{1.00,0.39,0.28}
\definecolor{turquoise1}{rgb}{0.00,0.96,1.00}
\definecolor{turquoise2}{rgb}{0.00,0.90,0.93}
\definecolor{turquoise3}{rgb}{0.00,0.77,0.80}
\definecolor{turquoise4}{rgb}{0.00,0.53,0.55}
\definecolor{turquoise}{rgb}{0.25,0.88,0.82}
\definecolor{violetred}{rgb}{0.82,0.13,0.56}
\definecolor{violet}{rgb}{0.93,0.51,0.93}
\definecolor{wheat1}{rgb}{1.00,0.91,0.73}
\definecolor{wheat2}{rgb}{0.93,0.85,0.68}
\definecolor{wheat3}{rgb}{0.80,0.73,0.59}
\definecolor{wheat4}{rgb}{0.55,0.49,0.40}
\definecolor{wheat}{rgb}{0.96,0.87,0.70}
\definecolor{whitesmoke}{rgb}{0.96,0.96,0.96}
\definecolor{white}{rgb}{1.00,1.00,1.00}
\definecolor{yellow1}{rgb}{1.00,1.00,0.00}
\definecolor{yellow2}{rgb}{0.93,0.93,0.00}
\definecolor{yellow3}{rgb}{0.80,0.80,0.00}
\definecolor{yellow4}{rgb}{0.55,0.55,0.00}
\definecolor{yellowgreen}{rgb}{0.60,0.80,0.20}
\definecolor{yellow}{rgb}{1.00,1.00,0.00}
 
\usepackage{ulem}
\definecolor{DarkGreen}{rgb}{0.00,0.39,0.00}
\definecolor{Turquoise}{RGB}{0,85,85}
\definecolor{TurquoiseLight}{RGB}{20,180,130}
\usepackage{tcolorbox}

\usepackage[color=TurquoiseLight,colorinlistoftodos]{todonotes}

\usepackage{amsmath,amsfonts}
\usepackage{amssymb}

\newcommand{\IBMmath}[0]{\textit{I\kern-1pt B\kern-1pt M}}
\shorttitle{ }\shortauthor{ }

\title{On the relation of secondary motions to turbulent large-scale motions}
\title{On the relation of secondary motions to large- and very-large-scale motions}
\author{
  Kay Schäfer\aff{1}\corresp{\email{kay.schaefer@uni-hamburg.de}} %, 
}

\affiliation{
\aff{1}Meteorology, Department of Earth System Sciences, University of Hamburg, Hamburg, Germany}
\begin{document}
\maketitle

% \tableofcontents
%\listoftodos
%#######################################################################
\noindent\rule{\textwidth}{0.5pt}\\[-3pt]
\begin{abstract}

The coexistence of secondary motions with large-scale and very-large-scale motions (LSMs/VLSMs) is investigated in turbulent open-channel flow over streamwise-aligned Gaussian ridges. 
Direct numerical simulations are performed at $Re_\tau = 540$ and $1000$ and the ridge geometry is varied by the ridge spacing $S$ and ridge height $h$.
Three main results are obtained.
First, the ridge height controls the strength of the secondary motions and thereby their coexistence with VLSMs. 
VLSMs are suppressed for $h \geqslant 0.05\delta$, with $\delta$ being the open-channel height,  and re-emerge at $h = 0.025\delta$, where the mean secondary motions are of similar strength to those over strip-type surfaces.
Previously reported differences in the coexistence behaviour over strip- and ridge-type surfaces thus arise from
the strength of the secondary motions rather than from the surface type.
Second, proper orthogonal decomposition and conditional averaging reveal instantaneous secondary motions that span approximately $2\delta$, fill the channel height irrespective of $S$ and $h$, and are stronger than the mean secondary motions. 
These instantaneous secondary motions are identified as LSMs that are spatially locked and amplified by the ridges.
Over dense ridges, they alternate between neighbouring ridges, such that the mean secondary motions appear weak and near-wall confined although VLSMs are suppressed. 
The ridge height thus sets the strength of the locked LSMs, whereas the ridge spacing sets where and for how long they persist. 
For weak locked LSMs, ridges spaced at their natural spanwise extent increase the VLSM energy, indicating that persistent locked LSMs concatenate into VLSMs. 
Third, long streamwise-coherent streaks are absent over dense ridges, while large-scale instantaneous secondary motions remain. 
This suggests that the formation of secondary motions is driven by intermittent cross-sectional motions rather than by long streamwise-coherent streaks, similar to the streak-less self-sustaining bursts proposed for near-wall turbulence by \cite{jimenezStreaksWallboundedTurbulence2022}.

\end{abstract}

%#######################################################################
\noindent\rule{\textwidth}{0.5pt}\\[-3pt]

% \begin{keywords}
%   Authors should not enter keywords on the manuscript, as these must be chosen by the author during the online submission process and will then be added during the typesetting process (see \href{https://www.cambridge.org/core/journals/journal-of-fluid-mechanics/information/list-of-keywords}{Keyword PDF} for the full list).  Other classifications will be added at the same time.
% \end{keywords}
%#######################################################################
\vspace{-22pt}
%#######################################################################
\section{Introduction}
\label{sec:introduction}
%#######################################################################

Turbulent flows over heterogeneous surfaces are ubiquitous in engineering and geophysical applications, where surface heterogeneity ranges from engineered patterns designed to reduce drag or enhance mixing to natural variations in roughness and topography. 
The effect of such heterogeneity on the exchange of momentum, heat and mass remains insufficiently understood, which for example limits its representation in weather and climate models \citep{bou-zeidPersistentChallengeSurface2020}. 
A mechanism by which surface heterogeneity modifies the flow far beyond the immediate vicinity of the surface is the formation of turbulent secondary motions of Prandtl's second kind, which are induced by spanwise heterogeneous surfaces and only occur in turbulent flows \citep{Bradshaw1987}. 
They manifest as counter-rotating vortices in the cross-sectional mean velocity field.
Secondary motions are accompanied by spanwise alternating low- and high-momentum pathways (LMPs and HMPs, respectively), with LMPs typically located in the upward and HMPs in the downward flow regions of the mean secondary motions
\citep{Barros2014,Willingham2014}.

Two types of spanwise heterogeneous surfaces are commonly distinguished, which are strip-type surfaces with spanwise variations in wall properties such as roughness \citep{Willingham2014,Chung_similartiy_2018,Neuhauser_SimulationTurbulentFlow_2022,frohnapfelFlowResistanceHeterogeneous2024,schmidtInfluenceReynoldsNumber2026}, and ridge-type surfaces with spanwise variations in wall elevation \citep{Goldstein1998secondary,Wang2006,Vanderwel2019,Stroh_SecondaryFlowHeat_2020,Zampiron_secondary_2020,Schafer_EffectSpanwiseHeterogeneous_2022,zhdanovInfluenceRidgeSpacing2024}.
The majority of studies have considered these two types separately, and only a few studies have addressed their combination \citep{Stroh_RearrangementSecondaryFlow_2020,Schafer_ModellingSpanwiseHeterogeneous_2022}. 
In both cases, the size and strength of the mean secondary motions are governed by the spanwise length scale $S$ of the heterogeneity, i.e.\ the ridge spacing or the strip width \citep{Vanderwel2015,Chung_similartiy_2018}. 
The secondary motions are strongest for $S = \mathcal{O}(\delta)$, with $\delta$ representing the open-channel height or boundary layer height, such that the counter-rotating vortices scale with $\delta$ and fill the entire cross-section. 
For $S \gg \delta$, they remain localised near the ridges or strip transitions and retain a size of order $\delta$, whereas for $S \ll \delta$ they weaken and scale with $S$ in size \citep{Wangsawijaya_effect_2020,Zampiron_secondary_2020}.

While the time-averaged secondary motions are well characterised, considerably less is known about the instantaneous secondary motions, i.e. the cross-sectional motions and momentum pathways in the instantaneous flow field that constitute the mean secondary motions.
The LMPs and HMPs exhibit pronounced lateral meandering \citep{kevin_cross-stream_2017,Vanderwel2019,Zampiron_secondary_2020,Wangsawijaya_effect_2020}, which is most intense for $S \approx \delta$ \citep{Wangsawijaya_effect_2020}. 
Over herringbone riblets, \cite{kevin_cross-stream_2017} suggested that the weak mean secondary motion, despite a modification of the entire boundary layer, results from the superposition of stronger instantaneous secondary motions.
The mean secondary motions thus do not necessarily represent the instantaneous secondary motions, such that both need to be distinguished.
Proper orthogonal decomposition (POD) of the flow over ridges confirmed the lateral meandering \citep{Vanderwel2019,Zampiron_secondary_2020}, but employed spanwise POD windows scaling with $S$, such that instantaneous structures wider than the ridge spacing could not be detected. 
How the mean secondary motions result from the instantaneous secondary motions, and to what extent both differ in size and strength, therefore remains open and is addressed in the present study.

Secondary motions occur in the same range of spanwise scales as the large-scale and very-large-scale motions (LSMs and VLSMs, respectively) that form naturally in wall-bounded turbulence over spanwise-homogeneous surfaces.
VLSMs were first identified in turbulent pipe flow \citep{kimVeryLargescaleMotion1999} and, termed superstructures, in turbulent boundary layers \citep{Hutchins_evidence_2007}, where they appear as meandering regions of high and low streamwise momentum extending over several $\delta$ in the streamwise direction. 
LSMs are shorter, with streamwise extents of approximately $1$--$4\delta$ \citep{kimVeryLargescaleMotion1999,montyComparisonTurbulentPipe2009}. 
\cite{hwangStatisticalStructureSelfsustaining2015} described both as parts of self-sustaining attached eddies, in which VLSMs correspond to the large-scale streaks carrying mainly streamwise turbulent kinetic energy, and LSMs to the associated cross-stream motions carrying wall-normal and spanwise turbulent kinetic energy. 
The formation mechanism of VLSMs remains under debate, and two hypotheses have been proposed.
The first attributes VLSMs to the streamwise concatenation of LSMs in the form of hairpin packets \citep{kimVeryLargescaleMotion1999}. 
The second considers large-scale streaks and cross-stream motions to sustain themselves autonomously through a self-sustaining process, in which the cross-stream motions amplify the streaks and the streaks regenerate the cross-stream motions through their instability and breakdown \citep{hwangSelfSustainedProcessLarge2010,floresHierarchyMinimalFlow2010,de_giovanetti_streak_2017}.
The concatenation hypothesis was supported by \citet{deshpandeEvidenceThatSuperstructures2023}, who showed that VLSMs in high-Reynolds-number boundary layers are composed of geometrically self-similar wall-attached motions. 
The central role of the streak instability in the self-sustaining process was questioned by \citet{jimenezStreaksWallboundedTurbulence2022}, who damped the longest streaks in turbulent channel flow and found that the near-wall bursts self-sustain without long streaks, suggesting that the streaks are a by-product rather than a driver of the turbulent cycle.

Despite their common range of spanwise scales, secondary motions differ from LSMs and VLSMs in two respects.
LSMs and VLSMs have no preferred spanwise position and are therefore masked by time averaging, whereas secondary motions are spatially locked to the surface heterogeneity and persist in the mean flow. 
Moreover, $\delta$-scale secondary motions fill the entire cross-section in both internal and external flows \citep{Vanderwel2015,Zampiron_secondary_2020,Wangsawijaya_effect_2020}, whereas VLSMs extend into the outer region only in internal flows and remain confined to the logarithmic and lower wake region in boundary layers \citep{montyComparisonTurbulentPipe2009}. 
Apart from these differences, secondary motions exhibit an analogous organisation, in which the LMPs and HMPs correspond to the large-scale streaks and the counter-rotating cross-sectional vortices to the cross-stream motions, and the meandering of the LMPs and HMPs resembles that of VLSMs \citep{kevin_cross-stream_2017,Vanderwel2019,Wangsawijaya_effect_2020}. 
Given the similarities between secondary motions and LSMs/VLSMs, a common formation mechanism has been suggested. 
The meandering of the LMPs was attributed to a lateral instability \citep{kevin_cross-stream_2017}, which over ridge-type surfaces was associated with a spanwise inflection point and termed secondary current instability (SCI), setting the characteristic meandering wavelength of the momentum pathways \citep{Zampiron_secondary_2020}.
\cite{Wangsawijaya_effect_2020} interpreted the secondary motions over strip-type surfaces as spanwise-locked VLSMs and suggested that the surface-induced large-scale streaks and rolls undergo a streak--vortex instability analogous to that over smooth walls \citep{hwangSelfSustainedProcessLarge2010,de_giovanetti_streak_2017}.
These hypotheses attribute the instantaneous dynamics of secondary motions to an instability of streamwise-elongated streaks. 
By suddenly removing a strip-type roughness pattern, \citet{andreolliTemporalDecaySecondary2025} showed, however, that the cross-sectional motion decays faster than the momentum pathways, which remain coherent over 15–20$\delta$ downstream, indicating that the pathways are generated by the cross-sectional motion transporting mean streamwise momentum.
In view of the findings of \cite{jimenezStreaksWallboundedTurbulence2022} and \cite{andreolliTemporalDecaySecondary2025}, it remains open whether the formation of secondary motions requires long streaks, or whether it is driven by the cross-sectional motion itself.

The coexistence of secondary motions and naturally occurring VLSMs has been examined to gain insight into the relation between both and thereby into the formation of secondary motions. 
While VLSMs coexist with the secondary motions over strip-type surfaces for all strip widths considered by \citet{Wangsawijaya_effect_2020}, their formation is suppressed over ridge-type surfaces for $S \lesssim 2\delta$, and for $S \leqslant 0.5\delta$ even in the outer region above the near-wall confined mean secondary motions \citep{Zampiron_secondary_2020}. 
\citet{medjnounReynoldsNumberEffects2026} confirmed the suppression of VLSMs in the LMPs of ridge-induced secondary motions in turbulent boundary layers, and observed a gradual recovery of VLSMs in the HMPs with increasing ridge spacing and Reynolds number. 
A similar interplay between ridge-induced secondary motions and naturally occurring large-scale convective rolls was observed in mixed convection flows \citep{Schafer_EffectSpanwiseHeterogeneous_2022}, where the convective rolls were considerably weakened by dense ridges and required larger buoyancy forces to form. 
These observations raise two questions. 
The first is why secondary motions over strip- and ridge-type surfaces interact differently with VLSMs and whether they constitute different coherent structures.
The second is how secondary motions over dense ridges suppress VLSMs beyond the region occupied by the mean secondary motions.

One possible explanation for the different behaviour over strip- and ridge-type surfaces is that the strength of the secondary motions, rather than the surface type itself, determines whether they coexist with VLSMs. 
The ridge height provides a means to test this hypothesis, since it controls the strength of the secondary motions at fixed ridge spacing, with the smooth wall recovered in the limit $h \to 0$. 
While the influence of the ridge spacing has been
studied extensively, the ridge height has not been varied systematically with respect to the coexistence of secondary motions and VLSMs, apart from preliminary results indicating that it governs the suppression of VLSMs \citep{Schaefer_TurbulentLargescaleStructures_2023}. 
The suppression of VLSMs above the near-wall confined mean secondary motions may be explained by instantaneous secondary motions that are larger and stronger than their mean, as suggested by \citet{kevin_cross-stream_2017}. 
Both hypotheses will be addressed and tested in the present study.
In addition, the relation between the instantaneous cross-sectional motion and the long streaks is examined to address whether the formation of secondary motions requires long streaks.

To this end, direct numerical simulations (DNS) of turbulent open-channel flow over streamwise-aligned Gaussian ridges are performed at $Re_\tau = 540$ and $1000$, similar to the experimental configuration of \citet{Zampiron_secondary_2020}, with systematic variation of the ridge spacing $S$ and the ridge height $h$. 
The flow configuration, the numerical method and the simulation database are described in \S\,\ref{sec:methodology}. 
The results in \S\,\ref{sec:results} start from the mean secondary motions (\S\,\ref{subsec:mean_secondary_motion}) and the instantaneous velocity fields (\S\,\ref{subsec:instantaneous_velocity}). 
The coexistence of secondary motions and VLSMs is then quantified by spectral analysis (\S\,\ref{subsec:spectral_analysis}). 
The instantaneous secondary motions are extracted by POD with a spanwise window independent of the ridge spacing (\S\,\ref{subsec:pod}), complemented by conditional averages based on the POD time coefficients (\S\,\ref{subsec:conditional_averages}), and compared with the large-scale structures over the smooth wall. 
The results are discussed in \S\,\ref{sec:discussion} and the conclusions are summarised in \S\,\ref{sec:conclusion}.

%#######################################################################
%#######################################################################
\section{Methodology}
\label{sec:methodology}
%#######################################################################
%#######################################################################
\begin{figure}
  \centering
  \includegraphics[width=\textwidth]{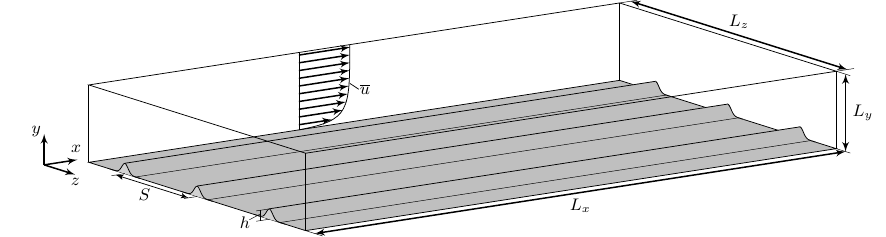}
  \caption{Sketch of the open-channel flow domain with streamwise-aligned Gaussian ridges. 
  }
  \label{fig:sektch_channel}
\end{figure}

The flow configuration for the present DNS is a fully developed turbulent open-channel flow with streamwise-aligned Gaussian ridges as depicted in figure \ref{fig:sektch_channel}. 
A constant pressure gradient is prescribed to maintain a constant friction Reynolds number $Re_\tau = u_\tau \delta_\textit{eff} / \nu$, with the friction velocity $u_\tau$, the effective open-channel height $\delta_\textit{eff}$ and the kinematic viscosity $\nu$.
The streamwise, wall-normal and spanwise directions are denoted by $x, y$ and $z$, respectively, or in index notation as $x_i$ ($i = 1, 2, 3$). 
The corresponding velocity components are $u, v$ and $w$, or $u_i$ in index notation.
Periodic boundary conditions (BCs) are applied in the horizontal directions, while in the vertical direction no-slip BCs are imposed at the surface wall and symmetry BCs ($v=0, \partial u/\partial y|_{y=\delta} = \partial w/\partial y|_{y=\delta} = 0$) at the top of the domain $y=\delta$, where $\delta$ is the open-channel height.

The continuity and Navier--Stokes equations %and temperature transport equation 
in index notation are 
\begin{gather} 
  \stepcounter{equation}
      \frac{\partial u_i}{\partial x_i} = 0,
      \qquad
      \frac{\partial u_i}{\partial t} + \frac{\partial u_i u_j}{\partial x_j}
          = - \frac{1}{\rho}\frac{\partial p}{\partial x_i}
            + \nu \frac{\partial^2 u_i}{\partial x_j \partial x_j }
            + \Pi \, \delta_{i1} % \frac{1}{\rho} P_x \delta_{i1}          
      \tag{\theequation\,a,b}%\\[5pt]
\end{gather} 
where $\rho$ is the density and $\Pi = u_\tau^2 / \delta_\textit{eff}$ is the prescribed constant pressure gradient.
The equations are solved by the open-source code Xcompact3d \citep{laizetHighorderCompactSchemes2009,bartholomewXcompact3DOpensourceFramework2020}, which is based on compact finite differences for the spatial discretisation and a third-order Runge--Kutta scheme for the temporal discretisation.
The representation of the ridges employs an immersed boundary method (IBM) that is based on polynomial reconstruction of the velocity fields inside the solid region \citep{Gautier_dns_2014}.
The code has previously been used to represent streamwise-aligned ridges in forced, mixed and natural convection \citep{Schafer_EffectSpanwiseHeterogeneous_2022}. 

The Gaussian ridge shape is identical to the one used in the mixed convection study by \cite{Schafer_EffectSpanwiseHeterogeneous_2022}, where the ridges are characterised by the ridge height $h$ and the spanwise width $\sigma$.
As in \cite{Schafer_EffectSpanwiseHeterogeneous_2022}, a value of $\sigma = 0.05\delta$ is used.  
The occupied cross-sectional area of a single ridge is $A_\textit{Gauss} = \sqrt{2\pi} h \sigma$, and thus the effective open-channel height is given by $\delta_\textit{eff} = \delta - \delta_\textit{melt}$, where the melt-down height is defined by $\delta_\textit{melt} = n_g A_\textit{Gauss} / L_z$, with $n_g$ the number of ridges in the domain \citep{Schafer_EffectSpanwiseHeterogeneous_2022}.

A triple decomposition \citep{Raupach_averaging_1982} is employed for the statistically spanwise heterogeneous flow fields, which yields for an arbitrary flow quantity $\phi$:
\begin{equation}
  \phi(x,y,z,t) = \langle \overline{\phi} \rangle (y) + \tilde{\phi}(y,z) + \phi''(x,y,z,t),
\end{equation}
where $\langle \overline{\phi} \rangle$ is the temporal, streamwise and spanwise mean, $\tilde{\phi}$ is the coherent fluctuation and $\phi''$ is the random fluctuation.
Angular brackets $\langle \cdot \rangle$ indicate spanwise averaging, while the overbar operator $\overline{(\cdot)}$ represents the temporal and streamwise mean.
The conventional turbulent fluctuation $\phi'$ is the sum of the coherent and random fluctuations
\begin{equation}
  \phi'(x,y,z,t) = \tilde{\phi}(y,z) + \phi''(x,y,z,t).
\end{equation}
In the presence of ridges, spatial averages are based on intrinsic averaging, which excludes grid points inside the ridges \citep{Schafer_EffectSpanwiseHeterogeneous_2022}.

\begin{table}
\begin{center}
\begin{tabular}{@{}l@{\hspace{2.5em}}l@{}}
%============================ LEFT: Re_tau = 540 =========================
\begin{tabular}[t]{cccccc}
\multicolumn{6}{c}{$Re_\tau = 540$}\\[3pt]
\hline\\[-8pt]
$S/\delta$ & $h/\delta$ & $\delta_\textit{eff}/\delta$ & $Re_{\tau,\textit{eff}}$ & $Re_b$ & $\Delta t_{tot}/t_b$ \\[3pt]
\hline\\[-6pt]
  $\infty$ & 0.000 & 1.000 & $540^\ast$     &  9850 & 1274 \\[3pt]
  %$\infty$ & 0.000 & 1.000 & $540^\dagger$  &  9860 &  467 \\[3pt]
      12.0 & 0.100 & 0.999 & $540^\ast$     &  9778 & 1796 \\[3pt]
       4.0 & 0.025 & 0.999 & $540^\ast$     &  9832 & 1488 \\
       4.0 & 0.050 & 0.998 & $540^\ast$     &  9775 & 1453 \\
       %4.0 & 0.050 & 0.998 & $540^\dagger$  &  9777 &  397 \\[2pt]
       4.0 & 0.100 & 0.997 & $540^\ast$     &  9618 & 1720 \\[3pt]
       %4.0 & 0.100 & 0.997 & $540^\dagger$  &  9640 &  333 \\[3pt]
       2.0 & 0.025 & 0.998 & $539$          &  9805 &  747 \\
       2.0 & 0.050 & 0.997 & $540$          &  9672 &  755 \\
       2.0 & 0.100 & 0.994 & $540^\ast$     &  9410 &  921 \\[3pt]
       1.0 & 0.025 & 0.997 & $540$          &  9784 &  755 \\
       1.0 & 0.050 & 0.994 & $539$          &  9641 &  762 \\
       1.0 & 0.100 & 0.987 & $540^\ast$     &  9232 &  804 \\[3pt]
       0.5 & 0.025 & 0.994 & $539$          &  9731 &  761 \\
       0.5 & 0.050 & 0.987 & $540$          &  9559 &  783 \\
       0.5 & 0.100 & 0.975 & $540^\ast$     &  9098 &  853 \\[3pt]
\hline
\end{tabular}
&
%============================ RIGHT: Re_tau = 1000 =======================
\begin{tabular}[t]{cccccc}
\multicolumn{6}{c}{$Re_\tau = 1000$}\\[3pt]
\hline\\[-8pt]
$S/\delta$ & $h/\delta$ & $\delta_\textit{eff}/\delta$ & $Re_{\tau,\textit{eff}}$ & $Re_b$ & $\Delta t_{tot}/t_b$ \\[3pt]
\hline\\[-6pt]
  $\infty$ & 0.000 & 1.000 & $1003$ & 19879 & 255 \\[3pt]
       4.0 & 0.100 & 0.997 & $1007$ & 19419 & 194 \\[3pt]
       2.0 & 0.025 & 0.998 & $1006$ & 19816 & 268 \\
       2.0 & 0.050 & 0.997 & $1007$ & 19488 & 298 \\
       2.0 & 0.100 & 0.994 & $1002$ & 18931 & 187 \\[3pt]
       1.0 & 0.100 & 0.987 & $1001$ & 18594 & 186 \\[3pt]
       0.5 & 0.025 & 0.994 & $1002$ & 19663 & 265 \\
       0.5 & 0.050 & 0.987 & $1002$ & 19396 & 268 \\
       0.5 & 0.100 & 0.975 &  $997$ & 18277 & 256 \\[3pt]
\hline
\end{tabular}
\end{tabular}
\caption{Simulation set-up, split by friction Reynolds number ($Re_\tau=540$ left, $Re_\tau=1000$ right).
The horizontal domain size is $L_x = 36\delta$ and $L_z = 12\delta$. 
The cases marked with $^\ast$ have been presented previously in \cite{Schaefer_TurbulentLargescaleStructures_2023}. %all remaining configurations are new to the present study.
The total time for the averaged statistics is indicated by $\Delta t_{tot}$ and $t_b$ is the bulk time unit.
The grid resolution for the $Re_\tau = 540$ cases is $N_x \times N_y \times N_z = 3072 \times 257\times 1152$ and for the $Re_\tau = 1000$ cases is $N_x \times N_y \times N_z = 5120 \times 385\times 2304$.
The grid is uniform in the horizontal directions and stretched in the wall-normal
direction. At $Re_\tau = 540$, the resolution is $\Delta x^+ = 6.3$, $\Delta z^+ = 5.6$
and $\Delta y^+ = 1$--$4.4$, and at $Re_\tau = 1000$ it is $\Delta x^+ = 7.0$,
$\Delta z^+ = 5.2$ and $\Delta y^+ = 1$--$6.8$.
}
\label{table:simulation_cases_split}
\end{center}
\end{table}

\vspace*{10pt}

The investigated simulation cases are listed in table \ref{table:simulation_cases_split}, ordered by decreasing $S$ and, for each $S$, increasing $h$. 
The smooth-wall configuration is denoted by $S=\infty$.
The effective friction Reynolds number $Re_{\tau,\textit{eff}}$ is computed from the wall-shear stress obtained by linearly extrapolating the total shear stress from the outer region to $y=\delta_{\textit{melt}}$ \citep{chan-braunForceTorqueActing2011}. 
Since $\Pi$ is prescribed, $Re_{\tau,\mathrm{eff}}$ recovers $Re_\tau$ in the statistically converged state, and its deviation from $Re_\tau$ therefore serves as a measure of statistical convergence.
The bulk Reynolds number is $Re_b = u_b \delta_\textit{eff}/\nu$, with the bulk velocity $u_b = 1/(L_z \delta_\textit{eff}) \int_0^{L_z} \int_{y_b(z)}^\delta \overline{u} \, \mathrm{d}y \, \mathrm{d}z$, where $y_b(z)$ is the local wall elevation.
The bulk time unit is denoted by $t_b=\delta/u_b$.
All simulations in table \ref{table:simulation_cases_split} use a horizontal domain size of $L_x \times L_z = 36\delta \times 12\delta$, which is sufficient to capture the spatial characteristics of VLSMs in turbulent open-channel flow \citep{bauerDirectNumericalSimulation2024}. 
The effect of the streamwise domain size is presented in Appendix \ref{app-subsec:influence_streamwise_domain} using a streamwise domain size of  $L_x = 108\delta$.

%#######################################################################
%#######################################################################
\section{Results}
\label{sec:results}
%#######################################################################

%#######################################################################
\subsection{Mean secondary motions}
\label{subsec:mean_secondary_motion}

%#######################################################################

\begin{figure}
    \centering
    \includegraphics[]{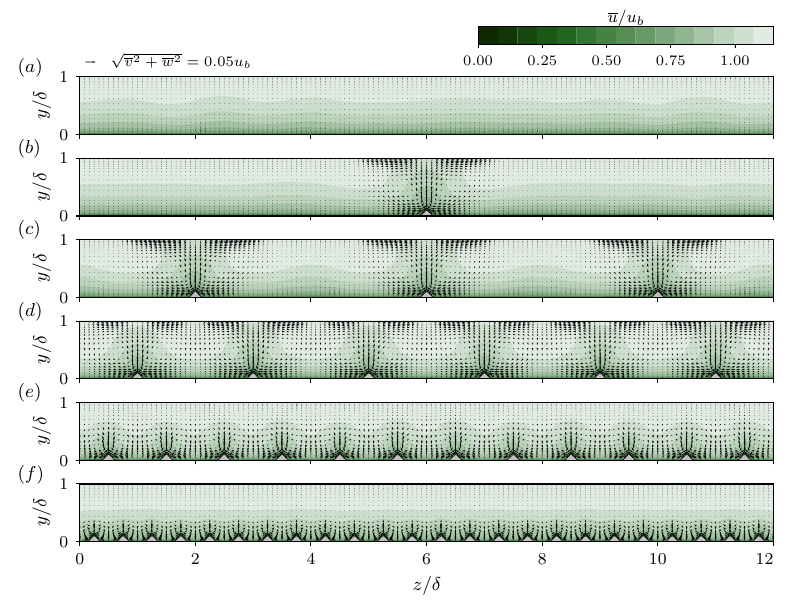}
    \caption{
    Mean streamwise velocity in the cross-sectional plane for $Re_\tau = 540$ and $h=0.1\delta$ with varying ridge spacing $S$. 
    The smooth-wall case is shown in $(a)$ and the Gaussian ridge cases are $S=12\delta$ $(b)$, $S=4\delta$ $(c)$, $S=2\delta$ $(d)$, $S=\delta$ $(e)$ and $S=0.5\delta$ $(f)$.
    The cross-sectional mean velocity components are illustrated by arrows and are scaled by $u_b$.\newline
  }
  \label{fig:mean2d_um_S_sweep_Retau540}
\end{figure}

First, we investigate the effect of the two ridge geometry parameters, ridge spacing $S$ and ridge height $h$, on the formation of mean secondary motions.
Figure \ref{fig:mean2d_um_S_sweep_Retau540} shows the effect of $S$ on the time- and streamwise-averaged velocity fields in the channel cross-section at $Re_\tau = 540$ with ridge height $h=0.1\delta$.
The mean secondary motion appears as counter-rotating vortices in the cross-sectional velocity components at the ridge location, which for wide ridge spacings $S\ge2\delta$ fills the entire open-channel height and displays a spanwise extent of about $2\delta$.
For smaller ridge spacings $S<2\delta$ the extent of the mean secondary motion scales with $S$, which is consistent with the findings of \cite{Zampiron_secondary_2020}, such that for the smallest ridge spacing case $S=0.5\delta$ the mean secondary motion is confined to the near-wall region, $y\lesssim 0.35\delta$.
The formation of secondary motions at the ridges is also accompanied by an upward bulging of the mean streamwise velocity, which is linked to the formation of LMPs.
The smooth-wall case in figure \ref{fig:mean2d_um_S_sweep_Retau540} $(a)$ shows a gentle bulging of the mean streamwise velocity at various locations, which is associated with the presence of VLSMs that have slow lateral dynamics and remain visible here due to the finite integration time. 
The largest ridge spacing case $S=12\delta$ in figure \ref{fig:mean2d_um_S_sweep_Retau540} $(b)$ shows a wide valley region that is not affected by the secondary motion, which allows for the formation of undisturbed VLSMs in the central valley region \citep[see][]{Schaefer_TurbulentLargescaleStructures_2023}.

%\

\begin{figure}
    \centering
    \includegraphics[]{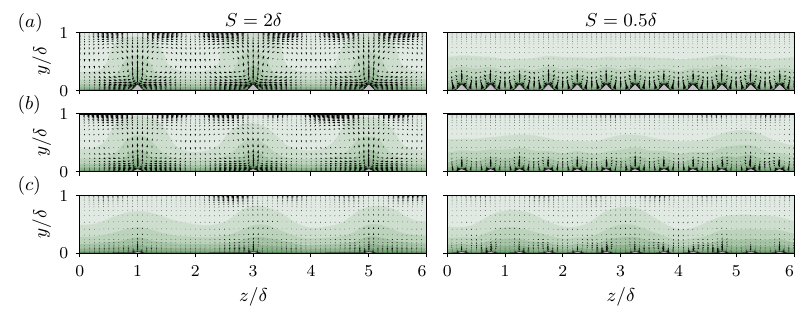}
    \caption{Influence of the ridge height $h$ on the mean streamwise velocity at $Re_\tau = 1000$ for case $S=2\delta$ (left column) and $S=0.5\delta$ (right column).
    Only one half of the cross-sectional domain is shown here.
    The ridge height variation shows $h=0.1\delta$ in (a), $h=0.05\delta$ in (b) and $h=0.025\delta$ in (c).
    The same colour map and arrow scaling as in figure \ref{fig:mean2d_um_S_sweep_Retau540} are used.
    }
    \label{fig:mean2d_um_hg_sweep_Retau1000}
\end{figure}

The effect of ridge height $h$ on the formation of mean secondary motion is shown in figure \ref{fig:mean2d_um_hg_sweep_Retau1000} for the ridge spacing cases $S=2\delta$ and $S=0.5\delta$ at $Re_\tau=1000$.
For both ridge spacings, reducing $h$ weakens the cross-sectional motion of the secondary motions, while the spatial extent remains qualitatively the same and is controlled by $S$.
These less intense mean secondary motions also result in a weaker bulging for the $S=2\delta$ case with decreasing $h$.
Interestingly, while the intensity of the cross-sectional motion also reduces for the $S=0.5\delta$ case with decreasing $h$, the mean streamwise velocity shows increasing bulging for smaller $h$, with a repetitive pattern in the spanwise direction every $2\delta$.
This selective bulging at certain ridges suggests that the ridges do not suppress the formation of VLSMs but rather fix them at certain spanwise positions.

%##################################################
% Strength of mean secondary motions
%##################################################

The strength of the mean secondary motions is quantified by the cross-sectionally averaged kinetic energy of the cross-sectional velocity components $u_{sm}$ \citep{schererRoleTurbulentLargescale2022}, which is defined by 
\begin{equation}
  u_{sm} = \left( \dfrac{1}{2\delta \,\delta_{\mathrm{eff}}} \int_{-\delta}^{\delta} \int_{0}^{\delta} (\tilde{v}^2 + \tilde{w}^2) \,  \mathrm{d}y \, \mathrm{d}z_c \right)^{1/2}. \label{equ:strength_mean_secondary_motion}
\end{equation}
The cross-sectional averages are based on phase-averaged velocity fields over $2\delta$-wide windows centred on a ridge, with $z_c$ the spanwise coordinate measured from the ridge centre.
The $2\delta$-wide window is chosen because it corresponds to the largest extent of the mean secondary motion as seen in figure \ref{fig:mean2d_um_S_sweep_Retau540} $(b,c)$ for the larger $S$ cases.

\begin{figure}
    \centering
    \includegraphics[]{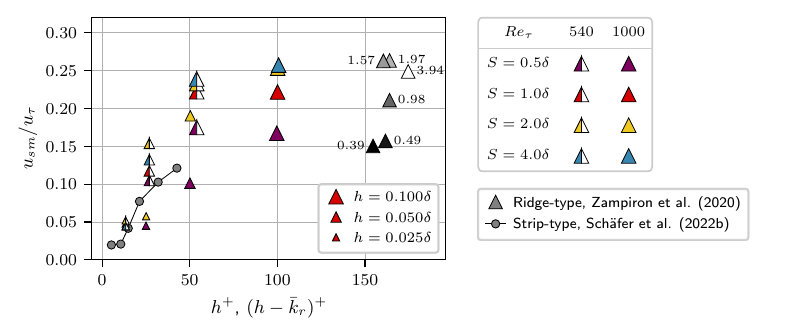}
    \caption{
    The strength of the mean secondary motions measured by $u_{sm}$ depending on the ridge height $h$ for different ridge spacings and Reynolds numbers. 
    The grey circles represent the strip-type and recessed roughness-strip cases from \cite{Schafer_ModellingSpanwiseHeterogeneous_2022}, with increasing relative height difference between the smooth-wall strip height $h$ and mean height of the roughness strip $\overline{k}_r$.
    The grey triangles indicate the ridge cases $h = 0.118\delta$ from \cite{Zampiron_secondary_2020}, where the different ridge spacings are coloured from black to white for low to high values of $S$ and the numbers next to the triangles provide the respective $S$ values.
    }
    \label{fig:strength_mean_secondary_motions}
\end{figure}

Figure \ref{fig:strength_mean_secondary_motions} presents the strength of secondary motions $u_{sm}$ for all ridge cases.
For a given pair of $Re_\tau$ and $h$ we see the previously noticed effect, that the cross-sectional motion of the secondary motions is the largest for $S\ge2\delta$, and for $S<2\delta$ the mean secondary motion scales with $S$ and weakens accordingly. 
The effect of ridge height is also consistent with the mean cross-sectional figures, where for a given pair of $Re_\tau$ and $S$, increasing the ridge height leads to an increase of $u_{sm}$.

The current ridge-type cases are compared to the DNS data of strip-type surfaces from \cite{Schafer_ModellingSpanwiseHeterogeneous_2022} at $Re_\tau = 500$, where the relative height difference between the smooth and rough wall strip $h-\bar{k}_r$ was varied systematically, with $h$ the height of the smooth-wall strip and $\bar{k}_r$ the mean height of the roughness strip. 
These strip-type cases also show that the strength of secondary motions increases with larger relative height differences, a behaviour which was attributed to the fact that these cases behave more like ridge-type surfaces \citep{Schafer_ModellingSpanwiseHeterogeneous_2022}.
The comparison with the ridge cases shows that the cases with the smallest ridge height, at a Reynolds number comparable to that of the strip-type cases, show a strength similar to that of the mean secondary motions of the pure strip-type cases.
This point will be important later for the discussion on the different coexistence behaviour of secondary motions and VLSMs seen for strip- and ridge-type surfaces.

For the experimental data of \cite{Zampiron_secondary_2020}, $u_{sm}$ was determined analogously to the present data by averaging the cross-sectional flow over a $2\delta$-wide window centred on the middle ridge of the open-channel flow.
Because the measurements do not resolve the flow between the ridges, the wall-normal integration covers only the region above the ridge crest.
Despite these differences, the experimental values at $Re_\tau = 1310$--$1480$ and $h = 0.118\delta$ fall within the range of $u_{sm}/u_\tau$ obtained for the present $h = 0.1\delta$ cases at $Re_\tau = 540$ and $1000$. 
Both data sets also exhibit the same dependence on the ridge spacing $S$, such that small $S$ yield weaker secondary motions, whereas the largest values of $u_{sm}$ occur for $S \geq 1.57\delta$, consistent with the present cases $S = 2\delta$ and $4\delta$. 
With $h^+ \approx 155$–$175$, the experiments extend the present parameter range to larger ridge heights in viscous units without a further increase of $u_{sm}/u_\tau$.

The overall picture indicates a saturation of the strength of secondary motions $u_{sm}/u_\tau$ for $h^+ \gtrsim 50$ for the largest ridge height $h = 0.1\delta$, irrespective of $S$. 
The data of \cite{Zampiron_secondary_2020} support this observation for larger $h^+$ values, such that the experimental values remain within the range of the present $h = 0.1\delta$ cases for all ridge spacings.
For the smaller ridge heights $h = 0.05\delta$ and $0.025\delta$, a similar saturation can be anticipated at smaller $h^+$ values, although higher Reynolds numbers would be required to confirm it.
A possible explanation for the saturation is that for $h^+ \gtrsim 50$ the ridge crest protrudes beyond the buffer layer, such that the spanwise heterogeneity of the Reynolds stresses driving the secondary motion is set by the ridge geometry rather than by viscous near-wall processes.

%#######################################################################
\subsection{Instantaneous velocity fields}
\label{subsec:instantaneous_velocity}
%#######################################################################

\begin{figure}
    \centering
    \includegraphics[]{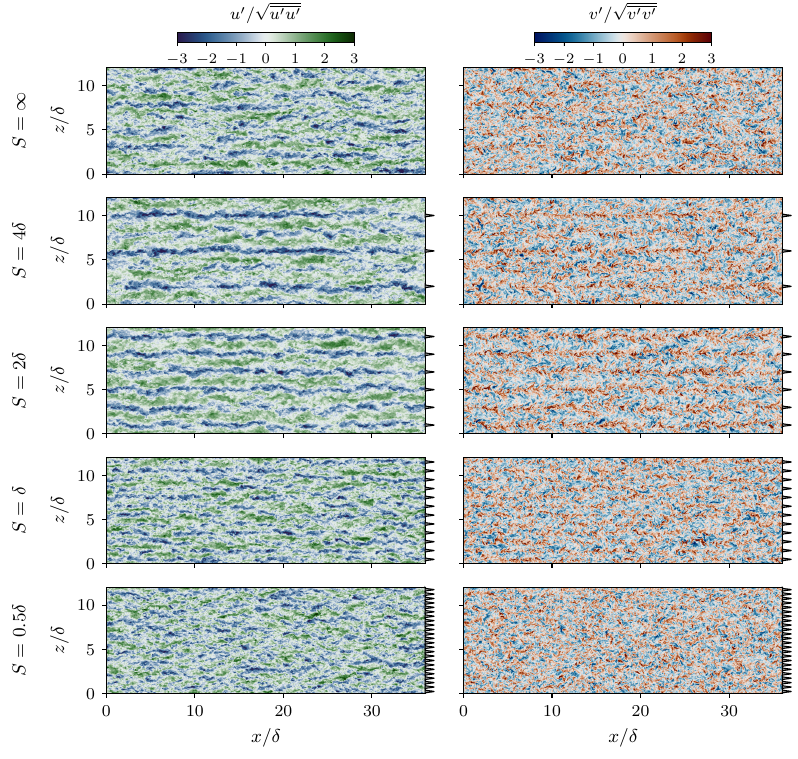}
    \caption{
    Instantaneous streamwise velocity fluctuations (left column) and wall-normal velocity fluctuations (right column) in a horizontal plane at $y = 0.5\delta$ for $h=0.1\delta$ and $Re_\tau = 1000$ for varying ridge spacing $S$.
    The smooth-wall case is shown in the top row and for the Gaussian ridge cases the sketched Gaussian ridges at the outer right figure frame indicate the spanwise position of the ridges.\newline 
    }
    \label{fig:yplanes_u_v_fluct_Retau1000_S_sweep}
\end{figure}

The occurrence of VLSMs in the presence of ridges is first examined using the instantaneous streamwise velocity fluctuation at the horizontal midplane ($y=0.5\delta$) for $Re_\tau = 1000$ and different ridge spacings $S$ shown in figure \ref{fig:yplanes_u_v_fluct_Retau1000_S_sweep}.
For the smooth-wall case ($S=\infty$), long meandering low- and high-momentum regions are observed, which are associated with VLSMs.
The larger ridge spacing cases $S\ge2\delta$ show pronounced low-momentum regions above the ridge location, and in the valley region long high-momentum regions occur that resemble the VLSMs in the smooth channel. 
Reducing the ridge spacing $S$ shortens the streamwise coherence of the LMPs above the ridges, and for $S=0.5\delta$ there are hardly any elongated momentum regions or long streaks visible. 
These results are consistent with the findings of \citet{Zampiron_secondary_2020}.

The organisation of the LMPs is also apparent in the instantaneous wall-normal velocity fluctuation field, shown in the right column of figure \ref{fig:yplanes_u_v_fluct_Retau1000_S_sweep}.
The meandering of the low-momentum region of the VLSMs in the smooth-wall case coincides with positive and strong $v'$. 
For the ridge cases $S\ge2\delta$ we see a similar organisation of $v'$ in the LMPs above the ridges, while for smaller $S$ this organisation weakens.
For $S=0.5\delta$ the distribution of $v'$ appears completely random and does not show similarities to the smooth-wall case, even though this midplane is above the region where the mean secondary motion influences the flow as shown in figure \ref{fig:mean2d_um_S_sweep_Retau540} $(f)$ and figure \ref{fig:mean2d_um_hg_sweep_Retau1000} $(a)$.
Similar observations can be made for the lower Reynolds number cases at $Re_\tau = 540$ \citep[not shown here, see][]{Schaefer_TurbulentLargescaleStructures_2023}, which are consistent with the experiments of \cite{Zampiron_secondary_2020} and illustrate that dense ridge spacings prevent the formation of VLSMs, which will be further quantified by the spectral analysis in \S\,\ref{subsec:spectral_analysis}.

\begin{figure}
  \centering
  \includegraphics[]{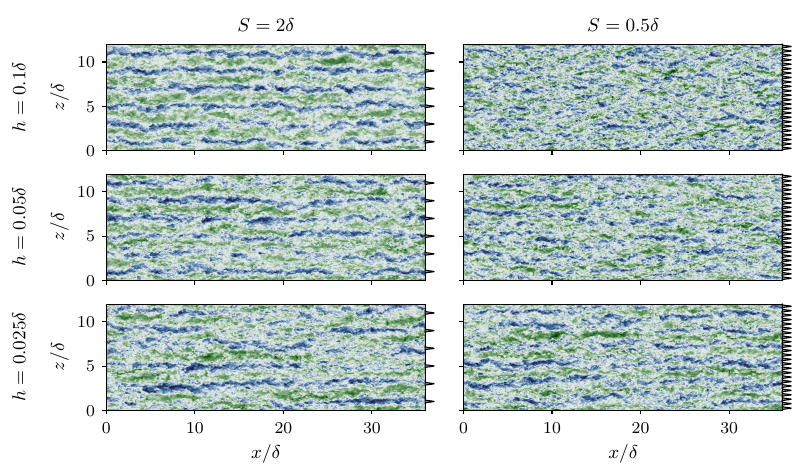}
  \caption{    
    Effect of ridge height $h$ on instantaneous streamwise velocity fluctuations in a horizontal plane at $y = 0.5\delta$ for $Re_\tau = 1000$ and $S=2\delta$ (left column) and $S=0.5\delta$ (right column).
    The same colour map as in figure \ref{fig:yplanes_u_v_fluct_Retau1000_S_sweep} is used.
  }
  \label{fig:yplanes_u_fluct_Retau1000_hg_sweep}
\end{figure}

The effect of the ridge height on the instantaneous flow organisation is shown for the two ridge spacings $S=2\delta$ and $S=0.5\delta$ at $Re_\tau=1000$ in figure \ref{fig:yplanes_u_fluct_Retau1000_hg_sweep}.
Reducing the ridge height for $S=2\delta$ reduces the streamwise coherence of the LMPs at the ridge location and the flow field gradually resembles the one shown for the smooth-wall case in figure \ref{fig:yplanes_u_v_fluct_Retau1000_S_sweep}.
The densest ridge spacing case $S=0.5\delta$ shows a different behaviour, where the reduction in $h$ allows for the gradual re-emergence of longer streamwise-coherent LMPs/HMPs and, for the smallest ridge height $h=0.025\delta$, the flow field eventually resembles that of the smooth-wall case.

%#######################################################################
\subsection{Spectral analysis}
\label{subsec:spectral_analysis}
%#######################################################################

\begin{figure}
  \centering
  \includegraphics[]{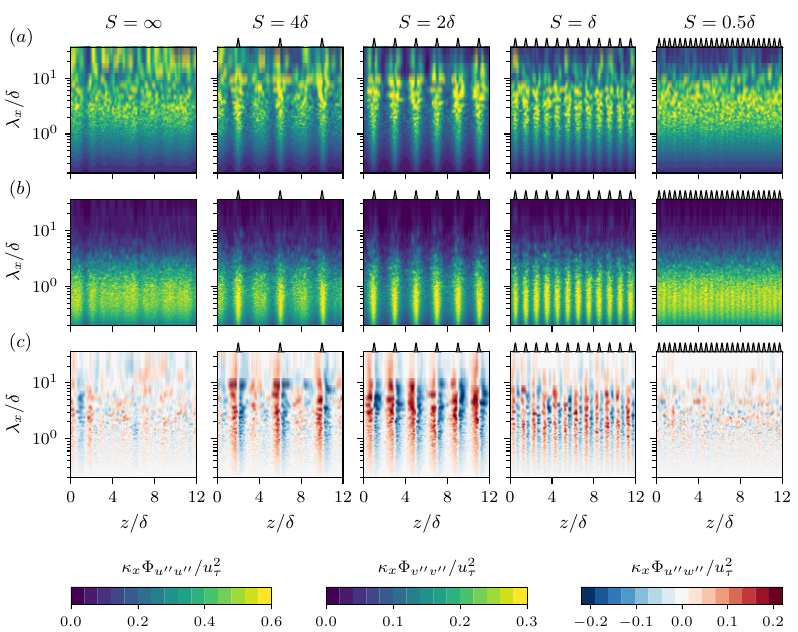}
  \caption{
    Pre-multiplied streamwise velocity spectra $\kappa_x \Phi_{u''u''} (\lambda_x,z)$ in $(a)$, pre-multiplied wall-normal velocity spectra $\kappa_x \Phi_{v''v''} (\lambda_x,z)$ in $(b)$ and velocity co-spectra $\kappa_x \Phi_{u''w''} (\lambda_x,z)$ in $(c)$ at the wall-normal location $y=0.5\delta$ for $Re_\tau=1000$ cases with varying ridge spacings $S$ and $h=0.1\delta$.
    The Gaussian ridges at the outer top subpanel frame indicate the spanwise position of the ridges.
    }
  \label{fig:spectra1d_S-sweep_uu_vv_uw}
\end{figure}

The occurrence of LSMs/VLSMs and secondary motions is further investigated by their spectral signature using the one-dimensional power spectral density $\Phi_{\phi'' \psi''}$ which is defined by 
\begin{equation}
  \langle \overline{\phi'' \psi''} \rangle (y,z) = \int_0^\infty \Phi_{\phi'' \psi''} (\kappa_x, y, z) \, \mathrm{d} \kappa_x,
\end{equation}
for two arbitrary flow variables $\phi$ and $\psi$, where $\kappa_x$ is the streamwise wavenumber, which is related to the streamwise wavelength $\lambda_x$ by $\kappa_x = 2\pi / \lambda_x$.
The spectra shown here are computed with at least $N_{pl,y}=2900$ horizontal planes with a temporal spacing of $\Delta t = 0.064t_b$. 

Figure \ref{fig:spectra1d_S-sweep_uu_vv_uw} $(a)$ shows the pre-multiplied streamwise velocity spectra $\kappa_x \Phi_{u''u''} (\lambda_x,z)$ at wall-normal location $y=0.5\delta$ for the smooth-wall case and different ridge spacing cases $S$ with $h = 0.1\delta$ for $Re_\tau=1000$. 
For the smooth-wall case ($S=\infty$), energy contributions of the VLSM ($\lambda_x \ge 18\delta$) are present across the entire channel width, consistent with previous experimental observations reporting streamwise wavelengths in the range of $10$--$40\delta$ \citep{Cameron_very-large-scale_2017, Zampiron_secondary_2020}. 
The individual peaks visible within the VLSM range result from the limited time integration of the simulation, whereby individual VLSMs reside longer at certain spanwise locations, thereby contributing to local spectral peaks. 
These spanwise-localised peaks are also detectable for the lower Reynolds number cases, for which the integration time is up to five times longer \citep[see][]{Schaefer_TurbulentLargescaleStructures_2023}.
The LSMs of the smooth-wall case are visible at wavelength ranges of $\lambda_x \approx 1$--$4\delta$, which is consistent with previous open-channel flow experiments and simulations \citep{Cameron_very-large-scale_2017,bauerDirectNumericalSimulation2024}.

The introduction of ridges progressively modifies this spectral picture. 
For $S=4\delta$, VLSMs persist in the valley regions and close to the ridges, but their energy contribution is visibly reduced at the ridge locations. 
For $S=2\delta$, this suppression is more pronounced and the valley regions are devoid of VLSM energy, which coincides with the downwash region of the secondary motions. At the ridge positions, an increased spectral signature at intermediate wavelengths of $\lambda_x \approx 2$--$12\delta$ is observed, flanked by low-energy regions on both sides. 
This spectral feature, associated with a lateral instability of the mean velocity profile and termed secondary current instability (SCI) by \cite{Zampiron_secondary_2020}, bridges the wavelength range between LSMs and VLSMs.
The self-similar spectral pattern associated with the secondary motions is recovered for all cases with $S \geq 2\delta$, but weakens for $S \leq \delta$ due to the mutual influence of neighbouring ridges. 
For $S=\delta$, VLSMs are further weakened, and for $S=0.5\delta$, no VLSM contributions are detectable, while the intermediate-wavelength signature remains strong. 
This is consistent with the finding by \cite{Zampiron_secondary_2020} that ridge-induced secondary motions suppress the formation of VLSMs for ridge spacings $S \leq 2\delta$. As demonstrated by the present data, for larger spacings VLSMs can coexist with the secondary motions. 
The limiting case of sufficiently wide valleys, in which undisturbed VLSM regions are possible, has been confirmed by a simulation at $Re_\tau = 540$ with $S = 12\delta$ \citep[not shown here; see][figure~4.8]{Schaefer_TurbulentLargescaleStructures_2023}.

The pre-multiplied wall-normal velocity spectra $\kappa_x \Phi_{v''v''} (\lambda_x,z)$ in figure \ref{fig:spectra1d_S-sweep_uu_vv_uw} $(b)$ show strong contributions for $\lambda_x \le 2\delta$ for the smooth-wall case. 
In the presence of ridges, the secondary motions intensify these contributions at the ridge locations, while reducing them on both sides of the ridge. 
This intensification persists down to $S=0.5\delta$, where it is present at every ridge location, highlighting that the updraft region of the secondary motion remains active up to wall-normal locations of $y=0.5\delta$.

Figure \ref{fig:spectra1d_S-sweep_uu_vv_uw} $(c)$ shows the pre-multiplied velocity co-spectra $\kappa_x \Phi_{u''w''} (\lambda_x,z)$. 
A non-vanishing $\overline{u''w''}$ arises from spanwise gradients of the mean streamwise velocity, which are largest at the spanwise inflection points of $\overline{u}$. 
The co-spectrum indicates where and at which streamwise scales these inflection points interact with the turbulence, and hence where lateral meandering of low-momentum pathways, as observed by \cite{Zampiron_secondary_2020}, can be sustained.
The opposite signs of the co-spectra on each side of a ridge are consistent with the spanwise $\bar{u}$ profile, such that positive $w''$ is associated with positive $u''$ on the left and negative $u''$ on the right side of the ridge. 
For $S \geq 2\delta$, co-spectral contributions are concentrated in the range $\lambda_x \approx 2$--$10\delta$, overlapping with the enhanced $\Phi_{u''u''}$ range associated with the SCI. 
For $S=\delta$, the co-spectral intensity decreases and the peak shifts towards smaller $\lambda_x$, while for $S=0.5\delta$ the co-spectrum resembles that of the smooth-wall case.
For the smooth wall, this contribution vanishes in the limit of long averaging times, as turbulent structures are free to travel across the full channel width. 
Nevertheless, a weak pattern similar to that observed near the ridges is found at $z \approx \delta$, which is attributed to a long-lasting VLSM that remains visible owing to the finite integration time of the present simulation.

\begin{figure}
  \centering
  \includegraphics[]{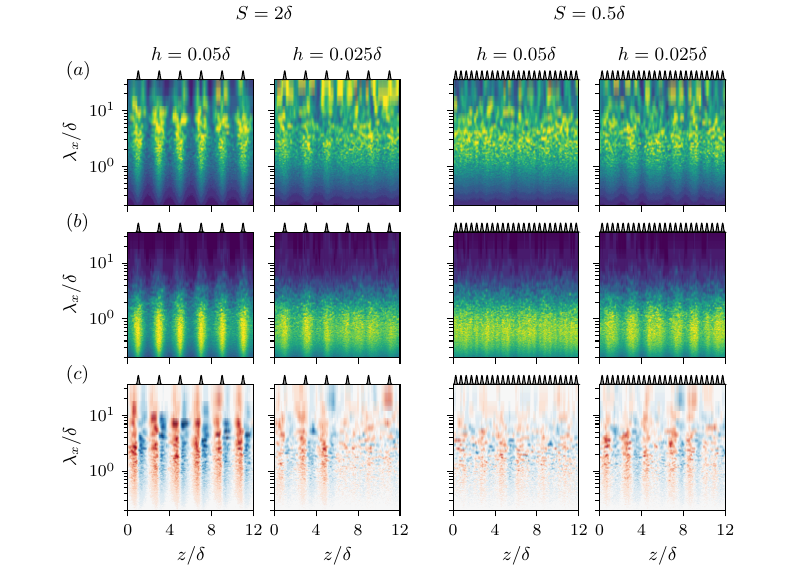}
  \caption{
    Pre-multiplied streamwise velocity spectra $\kappa_x \Phi_{u''u''} (\lambda_x,z)$ in $(a)$, pre-multiplied wall-normal velocity spectra $\kappa_x \Phi_{v''v''} (\lambda_x,z)$ in $(b)$ and velocity co-spectra $\kappa_x \Phi_{u''w''} (\lambda_x,z)$ in $(c)$ at the wall-normal location $y=0.5\delta$ for the two ridge spacing cases $S=2\delta$ and $S=0.5\delta$ at $Re_\tau = 1000$ with varying ridge height $h$.
    The Gaussian ridges at the outer top subpanel frame indicate the spanwise position of the ridges.
    The same colour map as in figure \ref{fig:spectra1d_S-sweep_uu_vv_uw} is used.
    }
  \label{fig:spectra1d_hg-sweep_uu_uw}
\end{figure}

Figure \ref{fig:spectra1d_hg-sweep_uu_uw} shows how the ridge height $h$ controls the coexistence between secondary motions and VLSMs for the two ridge spacing cases $S=2\delta$ and $S=0.5\delta$ at $Re_\tau=1000$.
Both cases show that the reduction in $h$ leads to a recovery of energy at large streamwise wavelengths $\lambda_x$, which for $h=0.025\delta$ results in the reappearance of VLSMs.
At the same time the strong secondary-motion signature with high contributions at intermediate $\lambda_x$ for the $h=0.1\delta$ cases weakens with decreasing $h$. 
Interestingly, the signature of the secondary current instability in the $u''w''$ spectra weakens for $S=2\delta$ with decreasing $h$, whereas it strengthens for $S=0.5\delta$, and the spanwise repeating pattern suggests a spanwise anchoring of VLSMs, which coincides with the upward bulging observed in the mean streamwise velocity field in figure \ref{fig:mean2d_um_hg_sweep_Retau1000}.
The strength of the VLSMs also differs between the two cases, where for $S=2\delta$ the VLSMs are comparable to or stronger than those over the smooth wall (figure \ref{fig:spectra1d_S-sweep_uu_vv_uw}), whereas for $S=0.5\delta$ they are weaker, indicating that densely spaced ridges still disturb their full formation.
These results show that the coexistence at the ridge location is controlled by the ridge height $h$, though there are qualitative differences depending on $S$.

%###########################################################
% spectra at ridge and valley centre 
%###########################################################

\begin{figure}
    \centering
    \includegraphics{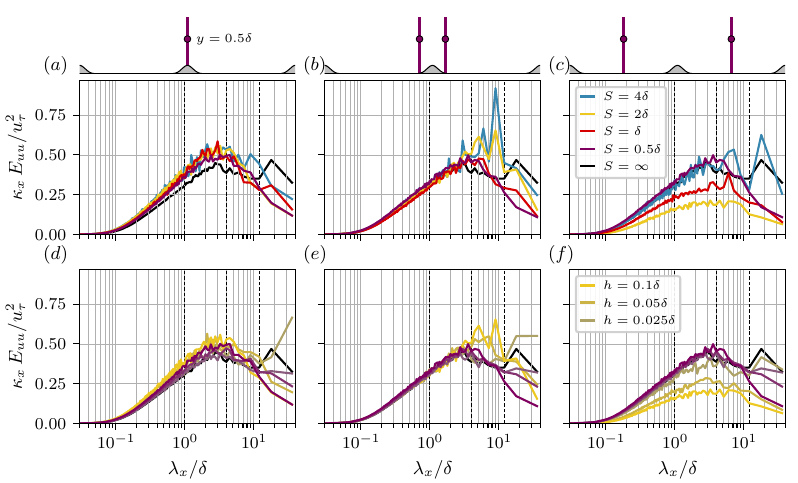}
    \caption{
      Pre-multiplied streamwise velocity spectra $\kappa_x \Phi_{u''u''} (\lambda_x)$ for $y=0.5\delta$ at different spanwise locations for the smooth-wall and ridge cases at $Re_\tau = 1000$.
      The panels $(a)$ and $(d)$ show the spectra at the ridge location ($z=z_c$), the panels $(b)$ and $(e)$ at the sides of the ridge location ($z=z_c \pm 0.25\delta$) and panels $(c)$ and $(f)$ in the centre of the valley ($z=z_c +S/2$). 
      Top row shows the effect of ridge spacing $S$ for $h=0.1\delta$ and the bottom row shows the effect of ridge height $h$ for the two ridge spacings $S=0.5\delta$ and $S=2\delta$.
      Phase averaging was applied to increase the statistical convergence of the results. 
      The vertical black dashed lines indicate the wavelength bands, which are used for the scale decomposition in figure \ref{fig:energy_contribution_VLSM_SCI_LSM_Retau1000}.
      %\textcolor{red}{Do not show $S=0.5\delta$ for the flank case, since this coincides with the valley centre region!}
      }
    \label{fig:spectra_uu_ridge_valley_Retau1000}
\end{figure}

To better quantify the suppression and recovery of VLSMs at specific spanwise positions,
figure~\ref{fig:spectra_uu_ridge_valley_Retau1000} compares the pre-multiplied streamwise velocity spectra
$\kappa_x \Phi_{u''u''}(\lambda_x)$ at $y = 0.5\delta$ at different spanwise locations (ridge location, flank of the ridges, centre of the valley), phase-averaged over all ridges, and compared with the spanwise-averaged smooth-wall
spectrum \citep{medjnounReynoldsNumberEffects2026}.

At the  ridge location the VLSMs are consistently suppressed across all $S$ cases with $h=0.1\delta$. 
Next to the ridge location ($z=z_c \pm 0.25\delta$) the VLSMs partly recover for $S=2\delta$ and $S=4\delta$, and in addition these cases show pronounced energy peaks in the wavelength range between LSMs and VLSMs ($\lambda_x \approx 5$--$10\delta$), which are associated with the secondary current instability mechanism of the LMPs above the ridges \citep{Zampiron_secondary_2020}.
In the valley region, case $S=4\delta$ shows a stronger peak at the VLSM wavelength than the smooth-wall case, which we attribute to the narrow valley centre region where VLSMs are constrained and thus last longer, while for the smooth-wall case the VLSMs can meander laterally over the full spanwise domain, such that the spanwise average diminishes their strength.
Interestingly, the partial recovery of the VLSMs at the flank of the ridges for $S=2\delta$ is not seen in the valley centre, which is attributed to the coinciding downdrafts of the neighbouring secondary motions in this area.
As noted by \cite{medjnounReynoldsNumberEffects2026} the energy content of the small scales ($\lambda_x<\delta$) collapses for all ridge cases, suggesting that secondary motions act on larger scales, whereas the smaller scales are merely redistributed.

The effect of the ridge height $h$ is shown in figure \ref{fig:spectra_uu_ridge_valley_Retau1000} $(d-f)$ for the ridge cases $S=0.5\delta$ and $S=2\delta$. 
Reducing the ridge height leads to a recovery of the energy at VLSM wavelengths in all shown spanwise locations.
For $S=2\delta$ and $h=0.025\delta$, the energy at the ridge location at the largest resolvable wavelength ($\lambda_x=36\delta$) is significantly larger than for the smooth-wall case, and at the flanks is even larger for $\lambda_x \ge 18\delta$, which suggests that for this specific ridge height $h$ and spacing $S$ the secondary motions can enhance the VLSM energy compared to the smooth-wall case. 
In the valley region, this increased energy at larger scales is not seen, which can be attributed to the coincidence of the downward motion of the secondary motions for this case. 
For the $S=0.5\delta$ cases, the energy at VLSM wavelengths is only recovered partly with decreasing $h$ and remains below the values of the smooth-wall case, indicating partial suppression of VLSMs by the secondary motions for dense ridges at the smallest ridge height. 

%###########################################################
% banded energy 
%###########################################################

\begin{figure}
    \centering
    \includegraphics[]{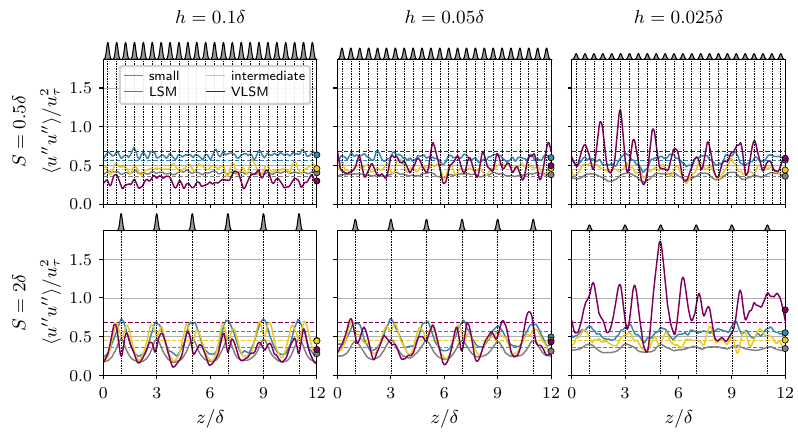}
    \caption{Spanwise distribution of the banded energy contribution for different turbulent scales (small scales, LSM, SCI and VLSM) and the effect of ridge height $h$ for two ridge spacings $S=0.5\delta$ (top row) and $S=2\delta$ (bottom row) at $Re_\tau=1000$.
    The dashed horizontal lines represent the mean values of the corresponding wavelength contribution of the smooth-wall case.
    The filled circles at the right subpanel frame indicate the spanwise mean values of the banded energy contribution of the shown case.
    }
    \label{fig:energy_contribution_VLSM_SCI_LSM_Retau1000}
\end{figure}

The energy contribution of the various coherent structures (LSM/VLSM and secondary motion) shown in the previous figures is quantified in more detail in figure \ref{fig:energy_contribution_VLSM_SCI_LSM_Retau1000} for $S=0.5\delta$ and $S=2\delta$ with varying ridge heights.
The energy contribution for the different structures is calculated as the integral of the streamwise velocity spectra over different wavelength ranges. 
For the LSMs the band $\delta < \lambda_x \le 4\delta$ is used, for the VLSMs $\lambda_x > 12\delta$ and in between the intermediate scales $4\delta < \lambda_x \le 12\delta$, which are associated with the observation of the SCI \citep{Zampiron_secondary_2020}.
For both ridge spacings at $h=0.1\delta$ the VLSM contribution is reduced by 55.9\% for $S=0.5\delta$ and 50.7\% for $S=2\delta$ in the spanwise mean compared to the smooth-wall mean. 
Case $S=2\delta$ additionally shows a clear spanwise modulation of the different contributions, where the VLSMs and SCIs feature two peaks adjacent to the ridge location, while the LSMs and small-scale motion peak directly at the ridge location.
The mean energy contribution of the intermediate scales does not change among the different ridge height and ridge spacing cases and remains within 2.8\% of the smooth-wall value, suggesting that secondary motions only reorganise the energy at these scales.
The LSM contribution for $S=0.5\delta$ and $h=0.1\delta$ is 12.0\% higher than the smooth-wall case and approaches the smooth-wall value with decreasing $h$, whereas for $S = 2\delta$ and $h = 0.1\delta$ it is 20.6\% lower and approaches the smooth-wall value with decreasing $h$.

Reducing the ridge height to $h = 0.05\delta$ slightly increases the VLSM contribution, more so for $S = 0.5\delta$ than for $S = 2\delta$, with spanwise-averaged values of 26.7\% and 36.0\% below the smooth-wall value, respectively. 
The contributions of the other wavelength bands remain largely unaffected. 
At $h = 0.025\delta$, the VLSM contribution increases substantially, such that for $S = 2\delta$ it exceeds the smooth-wall value by 25.1\%, whereas for $S = 0.5\delta$ it remains 12.6\% below it.
This suggests that a ridge spacing of $2\delta$, which corresponds to the natural spanwise size of LSMs, supports the formation of VLSMs, while denser ridges still slightly hinder their formation at the smallest ridge height.

%#######################################################################
\subsection{Proper orthogonal decomposition}
\label{subsec:pod}
%#######################################################################

While the results so far have shown that the coexistence behaviour at the ridge location between VLSMs and secondary motions is controlled by the ridge height $h$, an open question remains why ridges with $S=0.5\delta$ and $h=0.1\delta$ are able to suppress the formation of VLSMs also in the outer region (see figure \ref{fig:mean2d_um_S_sweep_Retau540} $(f)$ and figure \ref{fig:mean2d_um_hg_sweep_Retau1000} $(a)$), even though the mean secondary motion is near-wall confined. 
In contrast, for the dense strip-width cases of \cite{Wangsawijaya_effect_2020}, VLSMs were found to coexist with the near-wall confined secondary motions. 
To address this, the instantaneous flow structures are examined by means of the proper orthogonal decomposition (POD) to extract the most energetic flow structures \citep{lumleyStructureInhomogeneousTurbulent1967,Berkooz1993}.
The random velocity fluctuation field $\mathbf{u}''(y,z,t) = \left( u''(y,z,t), v''(y,z,t), w''(y,z,t) \right)^T$, which is interpolated on an equidistant grid in the wall-normal and spanwise direction, is decomposed by the POD as
\begin{equation}
    \mathbf{u}''(y,z,t) = \sum_{k=1}^{N} a_k(t)\,\boldsymbol{\Phi}_k(y,z),
\end{equation}
where $\boldsymbol{\Phi}_k(y,z)$ are the orthogonal spatial POD modes, $a_k(t)$ the corresponding POD time-coefficients and $N$ the total number of POD modes.
The POD modes are ordered by their energy in descending order, which is expressed by the eigenvalue $\lambda_k$, such that $\lambda_1 > \lambda_2 > \dots > \lambda_N$ holds.
The eigenvalues are computed by $\lambda_k = 1/ M \, \sum_{i=1}^M a_k^2(t_i) $, where $M$ is the total number of time snapshots or ensemble realisations used for the POD computation.
In the present study the number of spatial grid points exceeds the number of snapshots, such that the POD modes are efficiently computed with the method of snapshots \citep{sirovichTurbulenceDynamicsCoherent1987a}.

For the POD computation, cross-sectional planes at $N_{pl,x}$ different streamwise positions $x_i$ (here $N_{pl,x} = 9$), spaced $4\delta$ apart in the streamwise direction, are collected over $N_{t}$ time steps. 
These planes are further split into $N_{win}$ non-overlapping subwindows of width $L_{z,win}$, where for ridge cases these subwindows are centred at ridge locations \citep{Schaefer_TurbulentLargescaleStructures_2023}. 
The local spanwise coordinate of each subwindow is indicated by $z_c$, such that $q_{i,m}(y,z_c,t)$, with $q \in \{u,v,w \}$, represents the instantaneous velocity field of subwindow $m$ at the streamwise position $x_i$.
Here, we use a spanwise POD window size of $L_{z,win}=2\delta$, as tests for wider windows for $S=4\delta$ have shown that the largest energetic structures still exhibit a spanwise extent of about $2\delta$, which is shown in Appendix \ref{app-subsec:POD_spanwise_window_size}.
The total number of ensemble realisations $M = N_{pl,x} N_{win} N_{t}$ exceeds 50000 for all cases, and the time sampling interval $\Delta t$ lies between $0.13t_b$ and $0.27t_b$. 
The convergence of the POD results was examined for all three ridge heights of case $S=2\delta$ and $S=0.5\delta$ at $Re_\tau=1000$ by reducing the number of snapshots used for the POD computation to 50\% and 25\%. 
This results in a change in the eigenvalues of the first ten POD modes of less than $0.52\%$ for 50\% of the snapshots and less than $1.05\%$ for 25\% of the snapshots compared to the full set of snapshots.

\begin{figure}
    \centering
    \includegraphics[]{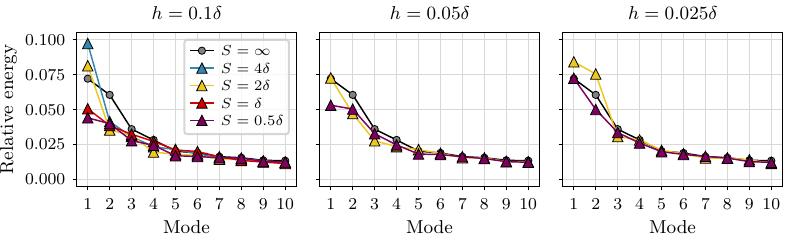}
    \caption{
    The relative energy contributions of the POD modes shown for different ridge height $h$ and ridge spacing $S$ cases at $Re_\tau = 1000$ for the first ten dominant POD modes.
    }
    \label{fig:pod_eigval_cum_energy_hg_sweep_Retau1000}
\end{figure}

Figure~\ref{fig:pod_eigval_cum_energy_hg_sweep_Retau1000} presents the relative energy contributions of the POD modes for different ridge spacings $S$ and ridge heights $h$ at $Re_\tau=1000$.
The comparison between different $S$ cases at $h=0.1\delta$ shows that the first POD mode is most energetic for the wider ridge spacing cases $S\ge2\delta$, with $9.7\%$ for $S=4\delta$ and $8.1\%$ for $S=2\delta$.
The denser ridge spacings exhibit a lower energy contribution in the first POD mode, $5.0\%$ for $S=\delta$ and $4.4\%$ for $S=0.5\delta$, which is even lower than the value for the smooth-wall case ($7.2\%$).
For the smooth-wall case, the first two eigenvalues lie in a similar range, indicating that these two POD modes represent a common structure.
The $S$-dependence of the relative energy observed for mode 1 weakens from mode 2 onward, such that the higher modes collapse onto a similar range across the $S$ cases.

The effect of the ridge height $h$ on the eigenvalues is examined for $S=2\delta$ and $S=0.5\delta$, which exhibit different trends with decreasing ridge height.
For $S=2\delta$, the dependence is non-monotonic, such that $\lambda_1$ decreases from $h=0.1\delta$ to $0.05\delta$, whereas at $h=0.025\delta$ both $\lambda_1$ and $\lambda_2$ increase, such that they exceed the smooth-wall values.
For $S=0.5\delta$, both eigenvalues increase slightly from $h=0.1\delta$ to $0.05\delta$. At $h=0.025\delta$, $\lambda_1$ increases further to the smooth-wall value, whereas $\lambda_2$ remains unchanged and below the smooth-wall value. The structures associated with these eigenvalues are discussed together with the spatial POD modes in the following.

\begin{figure}
    \centering
    \includegraphics[]{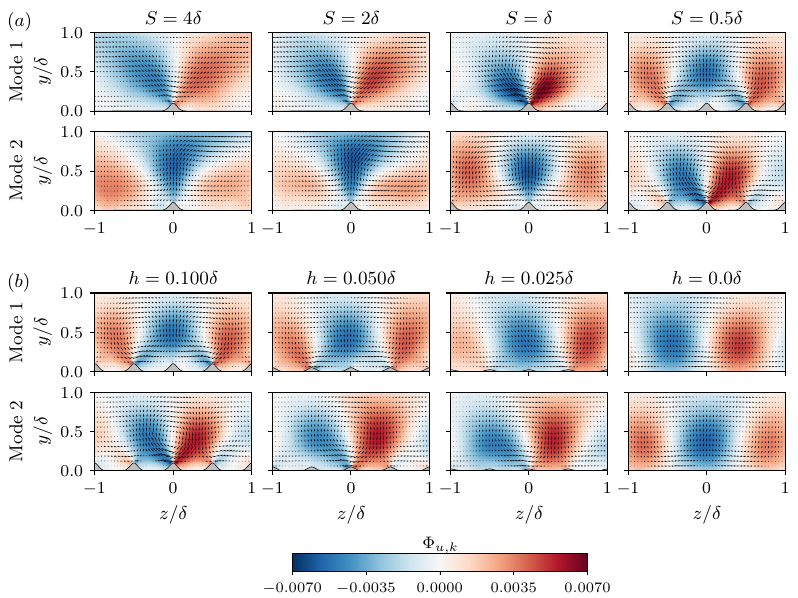}
    \caption{
    First two POD modes at $Re_\tau=1000$.
    (\textit{a}) Varying ridge spacing $S$ at fixed ridge height $h=0.1\delta$;
    (\textit{b}) varying ridge height $h$ at fixed ridge spacing $S=0.5\delta$, where the last column shows the smooth-wall case.
    The colour contours show the streamwise velocity component of the POD mode, while the arrows represent the cross-sectional velocity components for the POD mode.
    }
    \label{fig:pod_modes_S_hg_sweep_Retau1000}
\end{figure}

Figure \ref{fig:pod_modes_S_hg_sweep_Retau1000} shows the spatial pattern of the first two POD modes for different ridge spacings $S$ and ridge heights $h$ at $Re_\tau = 1000$.
As can be seen, all ridge cases exhibit large-scale coherent structures that fill the entire channel height, in contrast to the mean secondary motion for small $S$ shown in figure \ref{fig:mean2d_um_S_sweep_Retau540} and \ref{fig:mean2d_um_hg_sweep_Retau1000}.
The first POD mode for the $S\ge\delta$ cases at $h=0.1\delta$ represents a vortical structure centred at the central ridge location, which is associated with lateral shifts of the instantaneous secondary motions \citep{Vanderwel2019}.  
The second POD mode depicts large-scale counter-rotating vortices, resembling the mean secondary motions, representing strong instantaneous up- and downward motion above the central ridge. 
The results for the $S\ge\delta$ cases at $h=0.1\delta$ are consistent with the patterns found in previous studies \citep{Vanderwel2019,Zampiron_secondary_2020}.
However, for the denser ridge spacing case $S=0.5\delta$ we also observe large-scale structures resembling those found for larger ridge spacings. 
These structures for small $S$ were not found in the previous studies, since the spanwise POD window size was scaled by $L_{z,win} = S$, such that their full spatial extent was truncated.
The order of the first two POD modes is reversed for the densest ridge spacing $S=0.5\delta$. 
This reversal may result from neighbouring ridges diminishing the lateral shift dynamics of the secondary motions. 
Alternatively, laterally shifted structures at two adjacent ridges may combine into a common updraft between them at the central ridge, which enhances the counter-rotating vortex mode at the expense of the lateral-shift mode. 
The conditional averages in \S\,\ref{subsec:conditional_averages} support the latter, as they reveal events in which two adjacent ridges jointly sustain a single updraft.

The effect of the ridge height $h$ on the first two POD modes is shown for $S=0.5\delta$ in figure \ref{fig:pod_modes_S_hg_sweep_Retau1000} (b).
For all $h$ cases the second mode represents the lateral shift dynamics, which indicates that even for the smallest $h$ cases the influence of the neighbouring ridge is present such that the dominant mode remains the counter-rotating vortex pair in the first POD mode. 
Reducing the ridge height results in a less intense up- and downdraft region, such that the vortical patterns increasingly resemble those found for the smooth-wall case.
This is also found for the reduction of $h$ for $S=2\delta$ (not shown here), where for the smallest ridge height $0.025\delta$ the first POD mode represents counter-rotating vortices similar to the smooth-wall case.
The first two POD modes of the smooth-wall case represent counter-rotating vortices with similar eigenvalues that are phase-shifted by a quarter of their spanwise wavelength. 
Due to the spanwise translational invariance of the smooth-wall flow, such a mode pair represents the same coherent structure at arbitrary spanwise positions \citep{Berkooz1993,holmesTurbulenceCoherentStructures2012}.

The POD mode structures clarify the eigenvalue trends in figure \ref{fig:pod_eigval_cum_energy_hg_sweep_Retau1000}. 
For $S=0.5\delta$, the eigenvalue of the counter-rotating vortex mode recovers the smooth-wall value at $h=0.025\delta$, whereas that of the lateral-shift mode remains below the smooth-wall value for all ridge heights. 
The dense ridges thus do not weaken the instantaneous large-scale secondary motions relative to the smooth wall, but restrict their lateral shift dynamics. 
For $S=2\delta$ at $h=0.025\delta$, both leading modes represent counter-rotating vortices with the updraft displaced slightly to either side of the ridge, and their eigenvalues are nearly equal. 
This mode pair resembles that of the smooth-wall case, but remains localised at the ridge. 
At this ridge height, the ridge therefore no longer fixes the updraft at the ridge location, but keeps the structures in its vicinity, while their energy contribution exceeds that of the smooth wall.
This is consistent with the enhanced VLSM contribution at the ridge locations for this case (see figure \ref{fig:energy_contribution_VLSM_SCI_LSM_Retau1000}), indicating that the intensified large-scale structures at $h=0.025\delta$ coexist with VLSMs rather than suppressing them.

%#######################################################################
\subsubsection{Temporal dynamics of POD time coefficients}
\label{subsubsec:temporal_dynamics_time_coefficient}
%#######################################################################

\begin{figure}
    \centering
    \includegraphics[width=\textwidth]{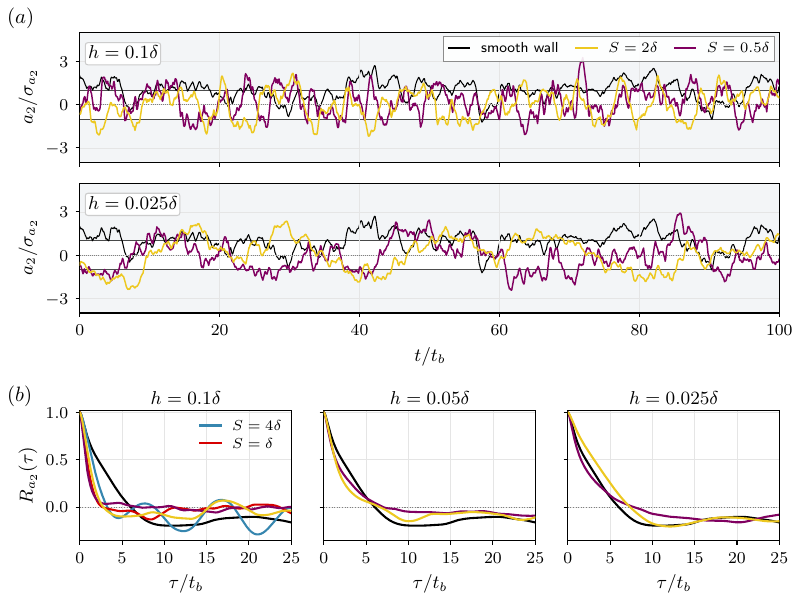}
    \caption{Time series of the POD time coefficient $a_2$ in $(a)$ for the ridge cases $S=2\delta$ and $S=0.5\delta$ at $h=0.1\delta$ and $h=0.025\delta$ and smooth-wall case at $Re_\tau=1000$, and the corresponding autocorrelation of $a_2$ for these cases in $(b)$.
    The figures in $(a)$ show a subset of the time series for a single subwindow for each case.
    For clarity, the cases $S=\delta$ and $S=4\delta$ with $h=0.1\delta$ are only shown for the autocorrelation in $(b)$.
    The light grey shaded area in $(a)$ illustrates the ranges $a_2 \ge \sigma_{a_2}$ and $a_2 \le -\sigma_{a_2}$, which highlight the time instances that are used for the conditional averages in \S\,\ref{subsec:conditional_averages}.
    }
    \label{fig:POD_time_coefficient_a2_time_series_auto_correlation}
\end{figure}

The presented POD modes for dense ridge spacings $S=0.5\delta$ suggest that strong instantaneous secondary motions with counter-rotating vortices appear either at the central or neighbouring ridges, depending on the sign of the corresponding time coefficient.
To compare the same coherent structures across the ridge cases, the first and second POD mode for the $S=0.5\delta$ cases are interchanged, such that the counter-rotating vortices are represented by the second POD mode.
To examine whether such an alternation occurs, the temporal behaviour of the second POD time coefficient $a_2$ is examined in figure \ref{fig:POD_time_coefficient_a2_time_series_auto_correlation} $(a)$ for $S=2\delta$ and $S=0.5\delta$ at $h=0.1\delta$ and $h=0.025\delta$, together with the smooth-wall case. 
This time coefficient exhibits frequent sign changes for both ridge cases at $h=0.1\delta$, on time scales of a few bulk time units, with $|a_2|$ frequently exceeding $\sigma_{a_2}$, where $\sigma_{a_2}$ is the standard deviation of $a_2$ computed across all subwindows.
For the smaller ridge height case $h=0.025\delta$, the time series of $a_2$ shows less frequent sign changes but extended periods of uniform sign lasting up to $20t_b$, resembling the characteristics of the smooth-wall case. 

To quantify this behaviour, the autocorrelation function of $a_2$ and the corresponding integral time scale are considered, which are defined as
\begin{equation}
  R_{a_k}(\tau)
  = \frac{\overline{a_k(t)\,a_k(t+\tau)}^{e}}{\sigma_{a_k}^2} ,
  \qquad
  T_{a_k} = \int_{0}^{\tau_0} R_{a_k}(\tau)\, \mathrm{d}\tau,
\end{equation}
where $\overline{(\cdot)}^{e}$ represents the ensemble average in time over all subwindows, $\sigma_{a_k}$ is the ensemble standard deviation of $a_k$ and $\tau_0$ is the first zero-crossing of $R_{a_k}$.

The autocorrelation of the second POD time coefficient is shown in figure \ref{fig:POD_time_coefficient_a2_time_series_auto_correlation} $(b)$.
For $h=0.1\delta$ the autocorrelation of the ridge cases decays considerably faster than for the smooth-wall case, with an integral time scale $T_{a_2}$ ranging from $0.96 t_b$ to $1.54 t_b$ for the ridge cases compared to $T_{a_2} = 2.46 t_b$ for the smooth-wall case. 
At the same time, a gentle $S$ dependence is observable for the ridge cases, where the decay rate decreases for larger $S$, which is also reflected in the integral time scale.
Reducing the ridge height leads to a reduction of the decay rate of the autocorrelation, and the ridge cases approach the behaviour of the smooth-wall case. 
The autocorrelation for $S=2\delta$ with $h=0.025\delta$ decays more slowly than for the smooth-wall case, with $T_{a_2} = 3.28 t_b$, which is consistent with the stronger VLSMs found for this case, which lead to longer temporal correlations.
For $S=0.5\delta$ with $h=0.025\delta$ the autocorrelation decays slightly faster than for the smooth-wall case with $T_{a_2} = 2.27t_b$, indicating that the dense ridge spacing configuration is still disturbing the formation of VLSMs. 

The integral time scale can be converted into a streamwise length scale of the structures represented by the second POD mode by invoking Taylor's hypothesis with the bulk velocity $u_b$ as convection velocity, which yields $L_{a_2} = u_b T_{a_2}$.
Since the bulk time unit is $t_b = \delta/u_b$, the values of $T_{a_2}/t_b$ reported above directly give $L_{a_2}/\delta$.
For the ridge cases with $h=0.1\delta$ the coherence length is thus of order $\delta$ and considerably shorter than for the smooth-wall case, whereas for $h = 0.025\delta$ the values of $L_{a_2}$ are comparable to or exceed the smooth-wall values. 
The cross-sectional structure captured by the second POD mode, which has LSM
dimensions in the spanwise and wall-normal directions
(see figure \ref{fig:pod_modes_S_hg_sweep_Retau1000}), therefore also possesses a streamwise coherence in line with the streamwise extent of LSMs \citep{montyComparisonTurbulentPipe2009}.

%#######################################################################
\subsection{Conditional averages}
\label{subsec:conditional_averages}
%#######################################################################

The POD modes presented in \S\,\ref{subsec:pod} show that similar dominant modes occur across all ridge spacings $S$ and ridge heights $h$, characterised by large-scale structures that span the entire channel height with a spanwise extent of $2\delta$.
By contrast, the mean flow fields in \S\,\ref{subsec:mean_secondary_motion} demonstrate that the mean secondary motion scales with $S$. 
This section examines conditional averages of the velocity fields based on the POD time coefficients, to investigate how these instantaneous large-scale structures organise spatially and 
give rise to smaller mean secondary motions for small $S$.
 
\begin{figure}
    \centering
    \includegraphics[]{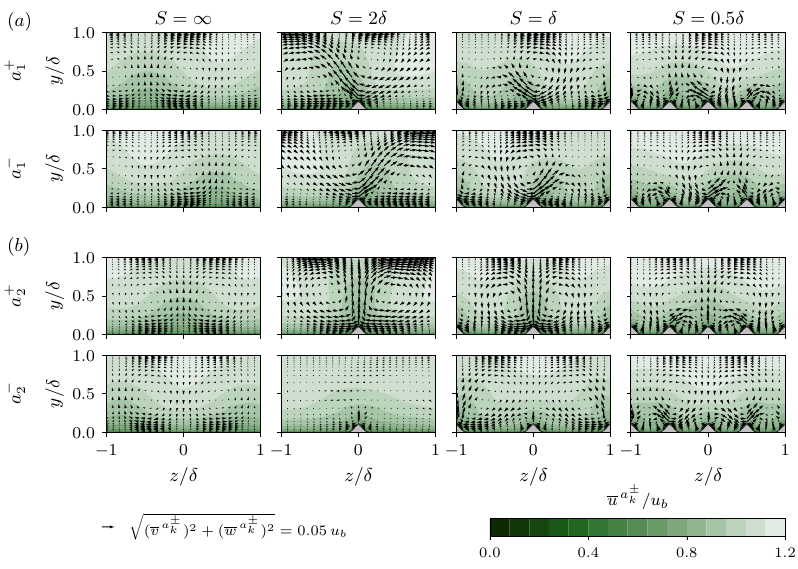}
    \caption{
    Conditionally averaged velocity fields based on the first POD time-coefficient $a_1$ (first two rows) and the second POD time-coefficient $a_2$  (last two rows) for the smooth-wall case (first column) and varying ridge spacing cases with $h=0.1\delta$ at $Re_\tau=1000$.
    }
    \label{fig:cond_averages_a1_a2_S_sweep_Retau1000}
\end{figure}

The conditional averages are based on the instantaneous velocity fields from the subwindows sampled for the POD computation in \S\,\ref{subsec:pod}.
For each subwindow $m$ at streamwise position $x_i$, two conditions are applied to the corresponding POD time coefficient $a_k^{i,m}$, expressed through the indicator function $I_k^{\pm \, (i,m)}$,
\begin{equation}
  I_k^{\pm \, (i,m)}(t) =
  \begin{cases}
    1, & \pm\, a_k^{i,m}(t) \geq \sigma_{a_k},\\[2pt]
    0, & \text{else},
  \end{cases}
  \label{eq:indicator}
  \qquad   
  N_k^{\pm} = \sum_{i=1}^{N_{pl,x}} \sum_{m=1}^{N_{win}} \sum_{t=1}^{N_t} I_k^{\pm \, (i,m)}(t),
\end{equation}
where $\sigma_{a_k}$ is the standard deviation of $a_k$ averaged over all subwindows, and $N_k^{\pm}$ is the total number of samples used for the respective condition. 
The standard-deviation threshold balances selecting sufficiently strong events against retaining sufficiently large ensembles (see figure \ref{fig:POD_time_coefficient_a2_time_series_auto_correlation} $(a)$).
For the first two POD time coefficients, the resulting fraction of retained time instances per condition ranges from $15.6\%$ to $18.4\%$ for all cases at $Re_\tau=1000$.
The conditional averages for the velocity field $q \in \{u,v,w\}$ are then obtained as the ensemble average over all subwindows, 
\begin{equation}
    \overline{q}_{}^{\,a_k^\pm}(y,z_c)
  = \frac{1}{N_k^{\pm}}
    \sum_{i=1}^{N_{pl,x}}\sum_{m=1}^{N_{win}}\sum_{t=1}^{N_t}
    I_k^{\pm\,(i,m)}(t)\; q_{i,m}(y,z_c,t).
\end{equation}
To ensure physical comparability across flow configurations, the signs of the POD modes are chosen such that a positive event corresponds for the first POD mode to a rightward-leaning vortical structure, and for the second POD mode to an upward motion over the central ridge (consistent with the sign convention used in figure \ref{fig:pod_modes_S_hg_sweep_Retau1000}).
Note that for the $S=0.5\delta$ cases the first and second POD mode are interchanged to allow for a direct comparison of similar coherent structures across the ridge cases.

\begin{figure}
    \centering
    \includegraphics[]{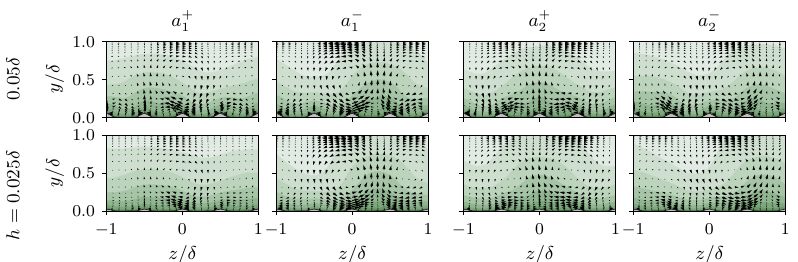}
    \caption{
    Conditionally averaged velocity fields for $S=0.5\delta$ with varying ridge height $h$ at $Re_\tau=1000$.
    The same colour bar and arrow size as in figure \ref{fig:cond_averages_a1_a2_S_sweep_Retau1000} are used.
    }
    \label{fig:cond_averages_a1_a2_hg_sweep_Retau1000}
\end{figure}

The conditional averages based on $a_1^\pm$ and $a_2^\pm$ are shown in figure \ref{fig:cond_averages_a1_a2_S_sweep_Retau1000} $(a)$ and $(b)$, respectively, at $Re_\tau=1000$ for the smooth-wall case and the ridge cases with varying ridge spacing $S$ at $h = 0.1\delta$.
For the smooth-wall case, all four conditions yield similar large-scale counter-rotating vortices that appear at different spanwise locations, indicating that they capture the same, spatially unanchored coherent structure. 
For the ridge cases, $a_1^\pm$ instead depicts a sidewards-leaning secondary-motion structure fixed at the central ridge, with the sign of the condition indicating a leftward- or rightward-leaning orientation.  
At the densest ridge-spacing case $S=0.5\delta$, this results in an updraft supported jointly by the central and the neighbouring ridge.  
The size and strength of these structures decrease slightly with decreasing $S$, but even at the densest spacing $S=0.5\delta$ they remain substantially larger and stronger than the total-mean secondary motion, and continue to span several ridges while driving a stronger updraft. 

The conditional averages based on $a_2^\pm$ in figure \ref{fig:cond_averages_a1_a2_S_sweep_Retau1000} $(b)$ show a strong secondary motion above the central ridge under $a_2^+$ and a weak one under $a_2^-$, relative to the total mean. 
For denser ridge spacings $S\le \delta$, the picture changes as the central-ridge motion remains weak under $a_2^-$, while a strong secondary motion appears at the neighbouring ridges.
These structures span the entire channel height with a spanwise extent of roughly $2\delta$, and therefore appear substantially larger than the total mean would suggest.
Together with the spatially unanchored behaviour established above for the smooth-wall case, this indicates that the ridges break the spanwise translational symmetry of the instantaneous large-scale structures, locking them to fixed spanwise locations rather than occurring freely in the domain.

Extending this comparison to the ridge-height variation, figure \ref{fig:cond_averages_a1_a2_hg_sweep_Retau1000} presents the conditional structures for $S=0.5\delta$ at $h=0.05\delta$ and $0.025\delta$. 
The structures retain the organisation observed for $h=0.1\delta$. 
The conditions on $a_1$ depict a sidewards-leaning secondary motion at the central ridge, which forms a strong updraft together with the adjacent ridge between the two. 
The condition $a_2^+$ yields counter-rotating vortices with a strong updraft above the central ridge, whereas under $a_2^-$ the secondary motion at the central ridge is weak and a sidewards-leaning structure appears at the neighbouring ridge, forming a common updraft with its next neighbour.
As for $h=0.1\delta$, these structures span the full channel height with a spanwise extent of approximately $2\delta$ at both smaller ridge heights, and therefore do not scale with the ridge spacing. 
With decreasing ridge height, the conditional structures increasingly resemble those of the smooth-wall case, consistent with the weaker lateral deflection of the flow at lower ridges and the reduced strength of the mean secondary motion (figure \ref{fig:strength_mean_secondary_motions}). 
In contrast to the smooth-wall case, however, the updrafts remain located either above a ridge or between two adjacent ridges, such that the ridges fix the spanwise position of the large-scale structures even at $h=0.025\delta$.

\begin{figure}
    \centering
    \includegraphics[]{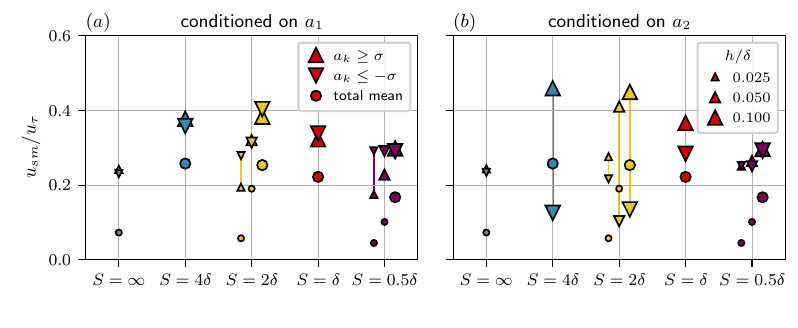}
    \caption{The strength of the conditionally averaged secondary motions based on the first POD time coefficient $a_1$ $(a)$ and second POD time coefficient $a_2$ $(b)$ for the different ridge cases at $Re_\tau=1000$. 
    The smooth-wall case is shown on the left of each panel by the grey symbols.
    Circles represent the strength of secondary motion based on the total mean shown in figure \ref{fig:strength_mean_secondary_motions}.
    }
    \label{fig:cond_averages_sm_strength_Retau1000}
\end{figure}

The strength of the conditionally averaged structures is evaluated following the formulation used for the mean secondary motion in equation \eqref{equ:strength_mean_secondary_motion}, applied here to the conditionally averaged cross-sectional motion, 
\begin{equation}
  u_{sm,a_k^{\pm}} = \left[ \frac{1}{2\delta \, \delta_{\mathrm{eff}} }
  \int_{-\delta}^{\delta} \int_{0}^{\delta}
  \left( \overline{v}_{}^{a_k^\pm\,2} + \overline{w}_{}^{a_k^\pm\,2} \right)
  \, \mathrm{d}y \, \mathrm{d}z_c \right]^{1/2} .
\end{equation}
Figure \ref{fig:cond_averages_sm_strength_Retau1000} shows the resulting strength for both conditions on $a_1$ and $a_2$, with the strength of the mean secondary motion from figure \ref{fig:strength_mean_secondary_motions} indicated by the circles for comparison.   
The strength under $a_1^+$ and $a_1^-$ is similar at $h=0.1\delta$, with both exceeding the mean, confirming the earlier observation that this mode reflects only the orientation of the secondary motion at the central ridge. 
For $a_2^\pm$ at the same $h$, the two conditions instead show a wide spread, with the positive condition substantially stronger than the mean. 
This spread is widest for $S\ge 2\delta$ and narrows with decreasing $S$, becoming comparable at $S=0.5\delta$, consistent with the alternation of the large-scale secondary motion between ridge locations at dense spacing.
Reducing the ridge height likewise reduces the strength of these conditional structures, which approach the values of the smooth-wall case. 

\begin{figure}
    \centering
    \includegraphics[]{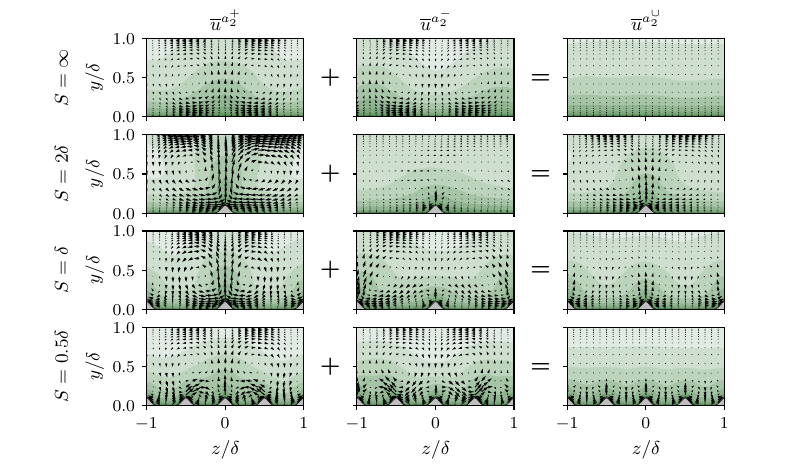}
    \caption{
    Comparison between conditionally averaged velocity fields based on the two conditions of the second time coefficient $a_2$ (first two columns) and their resulting sum (last column) for different $S$ cases with $h=0.1\delta$ at $Re_\tau=1000$.
    The smooth-wall case is shown in the first row.
    The colour bar and arrow size are the same as in figure \ref{fig:cond_averages_a1_a2_S_sweep_Retau1000} and \ref{fig:cond_averages_a1_a2_hg_sweep_Retau1000}.
    }
    \label{fig:cond_averages_a2_summation_S_sweep_Retau1000}
\end{figure}

To demonstrate how these conditional large-scale structures give rise to smaller mean secondary motions for small $S$, the combination of both conditions is considered, which is defined as
\begin{equation}
  \overline{q}^{a_k^\cup}
  = \frac{N_k^{+}\,\overline{q}^{a_k^+} + N_k^{-}\,\overline{q}^{a_k^-}}
         {N_k^{+} + N_k^{-}}.
\end{equation}
The result is shown in figure  \ref{fig:cond_averages_a2_summation_S_sweep_Retau1000} for the smooth-wall case and the ridge cases $S\le2\delta$ with $h=0.1\delta$ at $Re_\tau=1000$.
For the smooth-wall case, the two conditions cancel such that the large-scale structures are no longer apparent in the combined average. 
A similar result is obtained for the dense ridge spacing $S=0.5\delta$, where the large-scale structures are masked by the combined average and the resulting secondary motions instead scale with $S$, as observed for the total mean in figure \ref{fig:mean2d_um_S_sweep_Retau540} and \ref{fig:mean2d_um_hg_sweep_Retau1000}. 
Notably, this masking results not from temporal averaging of randomly positioned structures but from their alternation between fixed ridge positions.
For $S=2\delta$  and $S=\delta$, the combined conditions instead yield a weaker secondary motion resembling the total-mean structure. 
Taken together, these results demonstrate that the time mean masks the instantaneous large-scale dynamics of secondary motions, especially for dense ridge spacings.

%#######################################################################
\section{Discussion}
\label{sec:discussion}
%#######################################################################

Previous studies have shown contrasting coexistence behaviour between secondary motions and VLSMs over strip- and ridge-type surfaces. 
VLSMs coexist with secondary motions over strip-type surfaces \citep{Wangsawijaya_effect_2020}, whereas over ridge-type surfaces they are suppressed at the ridge location, in open-channel flows \citep{Zampiron_secondary_2020} as well as in turbulent boundary layers \citep{medjnounReynoldsNumberEffects2026}.
It was hypothesised in \S\,\ref{sec:introduction} that this difference arises from the strength of the secondary motions rather than from the surface type itself.

The present results support this hypothesis.
The ridge height controls the strength of the secondary motions and thereby determines whether they suppress the formation of VLSMs.
Ridges of height $h = 0.1\delta$, comparable to those of \cite{Zampiron_secondary_2020}, suppress VLSMs at the ridge location. 
Reducing the ridge height to $h = 0.05\delta$ only marginally increases the VLSM energy at the ridge flanks, which remains below the smooth-wall level. 
For the smallest ridge height ($h = 0.025\delta$), strong VLSMs emerge in the valleys and above the ridges at $S = 2\delta$, whereas at $S = 0.5\delta$ they occur only as isolated, locally intense events, such that the spanwise-averaged VLSM energy remains below the smooth-wall level.
The strength of the mean secondary motions at this ridge height is comparable to that over the strip-type surface of \cite{Schafer_ModellingSpanwiseHeterogeneous_2022}, whereas larger ridge heights yield strengths comparable to those over the recessed roughness strips of that study, which induce secondary motions characteristic of ridge-type surfaces \citep{hwang2018secondary}.
The coexistence behaviour is therefore governed by the strength of the secondary motions rather than by the surface type, which reconciles the previously reported differences in coexistence behaviour between strip- and ridge-type surfaces. 

The second question raised in \S\,\ref{sec:introduction} concerns the suppression of VLSMs over dense ridges, although the mean secondary motions are confined to the near-wall region. 
Over strip-type surfaces with $S \ll \delta$, VLSMs form above the mean secondary motions, whose wall-normal extent scales with $S$ \citep{Wangsawijaya_effect_2020}.
If the suppression were caused by the mean secondary motions, a similar coexistence would therefore be expected over ridges with $h = 0.1\delta$ and $S = 0.5\delta$, where the mean secondary motions are confined to $y \lesssim 0.35\delta$.
Instead, VLSMs are suppressed even above this region, consistent with the observations of \citet{Zampiron_secondary_2020} for dense ridge spacings and with \citet{medjnounReynoldsNumberEffects2026}, who report that secondary motions dominate at both ridge and valley locations for small $S$.
The secondary current instability (SCI), which \citet{Zampiron_secondary_2020} linked to the suppression of VLSMs at large ridge spacings, does not explain this behaviour. 
While its spectral signature at $\lambda_x \approx 2\delta$–$10\delta$ is present at the ridge locations for large ridge spacings, neither this signature nor a streamwise-coherent low-momentum pathway that could undergo such an instability is observed for $S = 0.5\delta$. 
Since the SCI is an instability of the mean flow and the mean secondary motions are confined to the near-wall region, the suppression of VLSMs at dense ridge spacings is not explained by the time-averaged flow field.

The suppression is instead explained by the instantaneous secondary motions, which, as hypothesised in \S\,\ref{sec:introduction}, are stronger than their mean and extend beyond the region occupied by the mean secondary motions. 
The dominant POD modes reveal large-scale secondary motions that span approximately $2\delta$ and fill the channel height for all ridge spacings and ridge heights. 
Conditional averages based on the first two POD time coefficients show that these structures are stronger than the mean secondary motions for all ridge spacings. 
Their strength is largest for $S \geqslant 2\delta$ and decreases with decreasing $S$, but still exceeds that of the smooth-wall structures for $S = 0.5\delta$. 
Since these instantaneous secondary motions extend over the full channel height also for $S = 0.5\delta$, they can account for the suppression of VLSMs above the region occupied by the mean secondary motions. 
Such structures were not detected for dense ridge spacings in previous studies, whose spanwise POD windows scaled with $S$ \citep{Vanderwel2019,Zampiron_secondary_2020}, whereas a fixed window of $2\delta$ is used here.

The discrepancy between the strong, full-height instantaneous secondary motions and the weak, near-wall confined mean secondary motions results from the alternation of the instantaneous structures between neighbouring ridges. 
Over dense ridges, the conditionally averaged secondary motions extend over several ridges, such that not every ridge can host an updraft simultaneously. 
At $S = 0.5\delta$, an updraft centred at one ridge prevents updrafts at the adjacent ridges on either side. 
In addition, the conditional averages show events in which two adjacent ridges jointly sustain a single updraft, which likewise prevents updrafts at their neighbouring ridges. 
The ridges thus fix the possible updraft locations, but not which ridge hosts the updraft at a given time. 
The short integral time scale of the corresponding POD time coefficients, of the order of one bulk time unit (see figure \ref{fig:POD_time_coefficient_a2_time_series_auto_correlation}), indicates that the updraft location changes rapidly between ridges. 
The superposition of these alternating structures yields a mean secondary motion that appears weak, near-wall confined and scales with $S$ (see figure \ref{fig:cond_averages_a2_summation_S_sweep_Retau1000}). 
The time mean thus masks the instantaneous secondary motions over dense ridges, as the weak mean reflects their alternation between neighbouring ridges rather than weak instantaneous structures.

The instantaneous secondary motions span approximately $2\delta$ and fill the channel height irrespective of the ridge spacing and height. 
Invoking Taylor’s hypothesis with $u_b$, the integral time scale of the second POD time coefficient corresponds to a streamwise coherence length of order $\delta$ for $h = 0.1\delta$, which increases to $2.3$–$3.3\delta$ for $h = 0.025\delta$ (see \S\,\ref{subsubsec:temporal_dynamics_time_coefficient}). 
The cross-sectional dimensions match those of LSMs in wall-bounded turbulence \citep{montyComparisonTurbulentPipe2009,hwangStatisticalStructureSelfsustaining2015}, and the streamwise coherence is comparable to or shorter than their streamwise extent. 
Structures of similar cross-sectional size and a coherence length of approximately $2.5\delta$ occur over the smooth wall, where they have no fixed spanwise position. 
The instantaneous secondary motions are therefore interpreted not as a distinct class of structures introduced by the surface, but as LSMs that are spatially locked and amplified by the ridges, hereafter referred to as locked LSMs.

Once the locked LSMs are weak enough for VLSMs to coexist with them, the ridge spacing modulates the VLSM energy through their persistence. 
With decreasing ridge height, the locked LSMs retain their large-scale structure, while their strength approaches that of the smooth-wall structures. 
At $h = 0.025\delta$, the spanwise-averaged VLSM energy exceeds the smooth-wall value by 25.1\% for $S = 2\delta$, whereas it is 12.6\% lower for $S = 0.5\delta$ (see figure \ref{fig:energy_contribution_VLSM_SCI_LSM_Retau1000}). 
The integral time scale of the POD time coefficients follows the same trend, which increases with decreasing $h$ and, at $h = 0.025\delta$, exceeds the smooth-wall value for $S = 2\delta$ while remaining slightly below it for $S = 0.5\delta$. 
Ridges spaced at the natural spanwise extent of the LSMs thus prolong the persistence of the locked LSMs, whereas dense ridges, each able to initiate a cross-sectional motion, shorten it. 
This suggests that weak and persistent locked LSMs contribute to the concatenation of LSMs into VLSMs \citep{deshpandeEvidenceThatSuperstructures2023}, while strong and short-lived locked LSMs at larger ridge heights disrupt it.

The locked-LSM interpretation also addresses the question raised in \S\,\ref{sec:introduction} of whether the formation of secondary motions requires long streaks.
\citet{Wangsawijaya_effect_2020} and \citet{wangsawijaya_investigation_2022} observed that instantaneous secondary motions over strip-type surfaces meander similarly to naturally occurring VLSMs and interpreted them as spanwise-locked VLSMs governed by a streak–vortex instability \citep{jeongCoherentStructuresWall1997,floresHierarchyMinimalFlow2010,de_giovanetti_streak_2017}, in which the cross-sectional motion results from the instability of the streaks. 
The present results instead identify the instantaneous secondary motions as locked LSMs, whose effect on VLSMs depends on their strength and persistence. 
Strong locked LSMs suppress VLSMs, whereas weak and persistent locked LSMs concatenate into VLSMs, as for $h = 0.025\delta$ and $S = 2\delta$. 
The spanwise-locked VLSMs observed by \citet{Wangsawijaya_effect_2020} over strip-type surfaces, which feature weaker secondary motions than ridge-type surfaces, are therefore likely a consequence of this concatenation rather than the primary structure of the secondary motions.
The identification of the secondary motions as spanwise-locked VLSMs may have been further complicated in turbulent boundary layers, where VLSMs are shorter than in internal flows \citep{montyComparisonTurbulentPipe2009}, and the spectral separation between LSMs and VLSMs is correspondingly smaller.

Whether strong locked LSMs form low-momentum pathways depends on the ridge spacing. 
Over dense ridges, large-scale cross-sectional motions occur without long streamwise-coherent streaks, indicating that the cross-sectional organisation does not require long streaks. 
A similar observation was made by \cite{jimenezStreaksWallboundedTurbulence2022}, who found that bursts self-sustain without long streaks.
These bursts are intermittent wall-normal ejections that resemble the cross-sectional motion considered here, and the streaks appear as their by-products rather than their drivers.
\citet{andreolliTemporalDecaySecondary2025} further found that the cross-sectional motion of secondary motions decays faster than the associated low-momentum pathway, such that the pathway outlives the cross-sectional motion.
Combined with the present short integral time scales of the POD time coefficients for $h = 0.1\delta$ and $S = 0.5\delta$, this suggests that the cross-sectional motion is re-initiated at a neighbouring ridge before a long streak can develop above an individual ridge. 
Over wider ridge spacings, the locked LSMs persist long enough above a ridge to form a low-momentum pathway by transporting mean streamwise momentum, consistent with the frequent observation of secondary motions coinciding with such pathways. 
These surface-fixed pathways are distinct from VLSMs, which remain suppressed at the ridge location for large ridge heights. 
The formation of secondary motions thus does not require long streaks and the low-momentum pathways are a consequence of persistent cross-sectional motion rather than its prerequisite.

%#######################################################################
\section{Conclusion}
\label{sec:conclusion}
%#######################################################################

The present study investigates the coexistence of turbulent secondary motions with large-scale and very-large-scale motions (LSMs/VLSMs) in turbulent open-channel flow over streamwise-aligned ridges, by varying the ridge spacing $S$ and the ridge height $h$ at two friction Reynolds numbers, $Re_\tau = 540$ and $1000$. 
The results show that the ridge height $h$ is the primary control parameter for the strength of the secondary motions and thereby for their coexistence with VLSMs. 
As $h$ is reduced, VLSMs remain suppressed for $h \geqslant 0.05\delta$ and re-emerge only at the smallest ridge height ($h = 0.025\delta$), for which the mean secondary motions are of similar strength to those over strip-type surfaces \citep{Schafer_ModellingSpanwiseHeterogeneous_2022}.
This indicates that the previously reported differences in the coexistence behaviour between strip- and ridge-type induced secondary motions with VLSMs \citep{Wangsawijaya_effect_2020,Zampiron_secondary_2020,medjnounReynoldsNumberEffects2026} arise from the strength of the secondary motions rather than
from the surface type.

The physical origin of this behaviour becomes apparent when distinguishing between mean and instantaneous secondary motions. 
Irrespective of $S$ and $h$, the instantaneous secondary motions are counter-rotating vortices with a spanwise extent of approximately $2\delta$ that fill the entire channel height, have a streamwise correlation length of $O(\delta)$ and have their upward flow centred above a ridge. 
Given these similarities to naturally occurring LSMs, the instantaneous secondary motions are identified as LSMs that are spatially locked and amplified by the ridges, hereafter referred to as locked LSMs. 
The mean secondary motions are thus the time-averaged signature of the locked LSMs.
Conditional averages show that the locked LSMs are considerably stronger than the mean secondary motions. 
For dense ridges ($S \leqslant \delta$), the locked LSMs alternate between neighbouring ridges, and their superposition yields mean secondary motions that are weak, near-wall confined and scale with $S$, which conceals the suppression of VLSMs by the strong and full-height instantaneous structures. 
The ridge height thus sets the strength of the locked LSMs and thereby determines whether VLSMs are suppressed, whereas the ridge spacing sets where and for how long the LSMs are locked, which governs the appearance of the mean secondary motions.

The results suggest that secondary motions are formed by the intermittent cross-sectional motion of locked LSMs rather than by long streamwise-coherent streaks.
For dense ridges, the large-scale cross-sectional motions are present although long streaks are absent, since the locked LSMs are re-initiated at a neighbouring ridge before such streaks can develop. 
Long streaks are thus a consequence of persistent cross-sectional motion rather
than its prerequisite, in line with the streak-less self-sustaining bursts in near-wall turbulence
\citep{jimenezStreaksWallboundedTurbulence2022}, such that the formation of secondary motions does not require an instability of long streaks \citep{Zampiron_secondary_2020,Wangsawijaya_effect_2020}.
For wider ridge spacings, the locked LSMs persist long enough above a ridge for low-momentum pathways to form, consistent with the commonly observed coincidence of secondary motions and such pathways. 
When the locked LSMs are weak ($h = 0.025\delta$), the ridge spacing controls the VLSM energy. 
Ridges spaced at the natural spanwise extent of the LSMs ($S = 2\delta$) prolong their persistence and increase the VLSM energy relative to the smooth wall, whereas dense ridges ($S = 0.5\delta$) shorten it and reduce the VLSM energy.
Spanwise-locked LSMs thus concatenate into VLSMs when they are weak and persistent, similar to the concatenation of LSMs into VLSMs over smooth walls \citep{deshpandeEvidenceThatSuperstructures2023}, which suggests that the spanwise-locked VLSMs observed over strip-type surfaces \citep{Wangsawijaya_effect_2020} result from a similar concatenation process.

The present results show that the coexistence of secondary motions with VLSMs is governed by the strength of the instantaneous secondary motions, i.e. the locked LSMs, which is set by the ridge height.
This implies that VLSMs over strip-type surfaces should be suppressed once a sufficient height difference between rough and smooth strips is introduced, which can be tested by varying this height difference as in \cite{Schafer_ModellingSpanwiseHeterogeneous_2022}, providing a continuous transition from strip-type to ridge-type behaviour. 
Varying the ridge height in the high-Reynolds-number boundary-layer experiments of \cite{medjnounReynoldsNumberEffects2026} would further test whether this role persists at larger scale separation. 
Finally, tracking the instantaneous secondary motions in time would allow the role of their persistence in the formation of low-momentum pathways and in the concatenation into VLSMs to be tested.

\vspace*{15pt}

%#######################################################################
%#######################################################################
%\backsection[Supplementary data]{\label{SupMat}Supplementary material and movies are available at \\https://doi.org/10.1017/jfm.2019...}

\backsection[Acknowledgements]{
The author is grateful to Davide Gatti and Juan Pedro Mellado for the insightful discussions regarding this work.
The author thanks Andrea Zampiron for providing the experimental data shown in figure~\ref{fig:strength_mean_secondary_motions}.
This work used resources of the Deutsches Klimarechenzentrum (DKRZ)
granted by its Scientific Steering Committee (WLA) under project ID 1276.
}

% \backsection[Note on AI use]{
% Claude Sonnet 5.5 and Claude Opus 5.5 (Anthropic) were used to improve language and readability and to suggest changes to sentence structure and organisation in parts of the manuscript. 
% All suggestions were reviewed and revised by the author, who takes full responsibility for the final text. 
% }

\backsection[Funding]{
KS acknowledges funding through the DFG Walter Benjamin Programme (project number 549151875).
}

\backsection[Declaration of interests]{The author reports no conflict of interest.}

\backsection[Data availability statement]{The data that support the findings of this study will be made publicly available upon publication of the manuscript.}

\backsection[Author ORCIDs]{\\[2pt]
K. Schäfer,  \href{https://orcid.org/0000-0002-1704-8233}{https://orcid.org/0000-0002-1704-8233}; \\[2pt]
}

% \backsection[Author contributions]{ KS: Conceptualisation, Methodology, Investigation, Validation, Visualisation, Data Curation, Project Administration, Formal Analysis, Supervision, Writing, Funding Acquisition}

%#######################################################################
%#######################################################################
% Appendix
%#######################################################################
\appendix

%###############################################
\section{Influence of streamwise domain size}
\label{app-subsec:influence_streamwise_domain}
%###############################################

\begin{table}
\begin{center}
\begin{tabular}{@{}l@{\hspace{2.5em}}l@{}}
%============================ LEFT: Re_tau = 540 =========================
\begin{tabular}[t]{cccccc}
\multicolumn{6}{c}{$Re_\tau = 540$}\\[3pt]
\hline\\[-8pt]
$S/\delta$ & $h/\delta$ & $\delta_\textit{eff}/\delta$ & $Re_{\tau,\textit{eff}}$ & $Re_b$ & $\Delta t_{tot}/t_b$ \\[3pt]
\hline\\[-6pt]
  $\infty$ & 0.000 & 1.000 & $540$  &  9860 &  467 \\
       4.0 & 0.050 & 0.998 & $540$  &  9777 &  397 \\
       4.0 & 0.100 & 0.997 & $540$  &  9640 &  333 \\
\hline
\end{tabular}
\end{tabular}
\caption{Simulation set-up of the long streamwise domain configurations at $Re_\tau=540$.
The streamwise and spanwise domain sizes are $L_x = 108\delta$ and $L_z = 12\delta$.
The simulations use a grid resolution of $N_x \times N_y \times N_z = 7168 \times 257\times 1152$.
}
\label{table:simulation_cases_long_domain}
\end{center}
\end{table}

\begin{figure}
    \centering
    \includegraphics[]{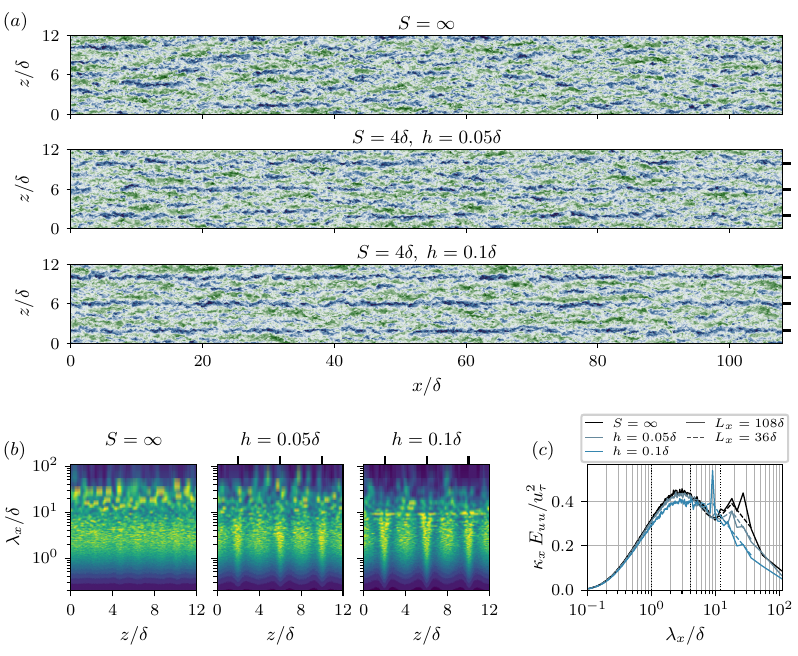}
    \caption{
    Effect of streamwise domain size on turbulent large-scale structures.
    In $(a)$ the instantaneous streamwise velocity fluctuation is shown in a horizontal plane at $y=0.5\delta$ for the long domain cases with different ridge heights at $Re_\tau = 540$.
    The pre-multiplied streamwise velocity spectra $\kappa_x \Phi_{u''u''} (\lambda_x,z)$ of the same cases at the wall-normal location $y=0.5\delta$ are shown in $(b)$.
    The spanwise-averaged pre-multiplied streamwise velocity spectra $\kappa_x \Phi_{u''u''} (\lambda_x)$ are presented in $(c)$ with the respective reference cases with streamwise domain length $L_x=36\delta$.
    The colour bars for the velocity fluctuations and for the velocity spectra are the same as shown in figure \ref{fig:yplanes_u_v_fluct_Retau1000_S_sweep} and figure \ref{fig:spectra1d_S-sweep_uu_vv_uw}.
    }
    \label{fig-app:yplanes_spectra}
\end{figure}

To investigate the effect of the streamwise domain size on the formation and representation of VLSMs and secondary motions additional simulations with $L_x=108\delta$ are performed at $Re_\tau = 540$. 
A smooth-wall case and two ridge configurations with $S=4\delta$ and different ridge heights ($h = 0.1\delta, 0.05\delta$) are presented, and their set-up is listed in table \ref{table:simulation_cases_long_domain}.
The streamwise velocity fluctuation at the horizontal midplane is shown for the three cases in figure \ref{fig-app:yplanes_spectra} $(a)$.
The smooth-wall case shows, as in figure \ref{fig:yplanes_u_v_fluct_Retau1000_S_sweep}, long coherent streaks that represent the VLSMs and occur throughout the spanwise domain.
For the ridge cases, long coherent low-momentum streaks can be seen at the ridge positions, which are more pronounced for the higher ridge height case $h=0.1\delta$ and are more coherent, although they are still intermittently interrupted. 

The spectral signature of the VLSMs and secondary motion is shown in figure \ref{fig-app:yplanes_spectra} $(b)$ using the pre-multiplied streamwise velocity spectra at the horizontal midplane.
For the long-domain cases the VLSMs show similar energy contribution in the range of $\lambda_x \approx 10–40\delta$ as for the $L_x=36\delta$ case at $Re_\tau=1000$ in figure \ref{fig:spectra1d_S-sweep_uu_vv_uw}, with isolated peaks reaching up to $50\delta$.
For the ridge cases, as for the short-domain cases, it is evident that the ridges attenuate the energy contribution of the VLSMs at the ridge location, and suppress it for $h = 0.1\delta$, while the energy contribution of the secondary motions occurs in the intermediate wavelength range of $\lambda_x \approx 2–12 \delta$.
Figure \ref{fig-app:yplanes_spectra} $(c)$ shows the comparison of the spanwise-averaged pre-multiplied streamwise velocity spectra for the long-domain cases and the corresponding short-domain cases.
The comparison reveals similar spectral profiles with comparable energy contributions for the secondary motions and VLSMs.
The longer domain results in a finer wavenumber resolution ($\Delta \kappa_x = 2\pi/L_x$), such that for the smooth-wall case, two peaks of the VLSMs now appear in the $\lambda_x = 18–30\delta$ range, which are not detectable for the short domain cases.
For the ridge cases, the longer domain cases exhibit similar energy contributions and no additional energy contributions appear for wavelengths $\lambda_x > 36\delta$.
Only for case $h = 0.1\delta$ is the secondary-motion peak at $\lambda_x=9\delta$ sharper for the long domain case, while this energy is spread across neighbouring wavelengths for the short domain case. 
Overall, the chosen domain size of $L_x = 36\delta$ is sufficient for investigating the interaction between VLSMs and secondary motions in this study.

%#######################################################################
\section{Secondary current instability}
\label{app-subsec:secondary_current_instability}
%#######################################################################

\begin{figure}
    \centering
    \includegraphics[]{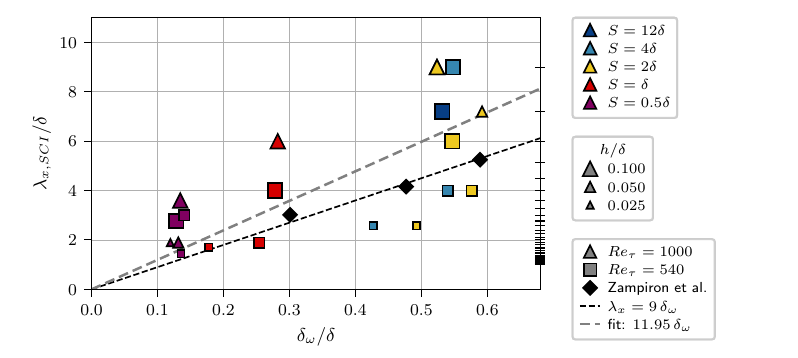}
    \caption{
    Streamwise wavelength of the secondary current instability $\lambda_{x,SCI}$ versus vorticity thickness $\delta_\omega$ for the present ridge cases, with both quantities evaluated at the cell centre of the mean secondary motion. 
    The colours indicate the ridge spacing $S$ and the symbol sizes the ridge height $h$, while triangles and squares distinguish between $Re_\tau = 1000$ and $Re_\tau = 540$. 
    For comparison, the ridge cases $\delta \leqslant S \leqslant 2\delta$ of \citet{Zampiron_secondary_2020} are included as diamonds together with their fitted relation $\lambda_{x,SCI} = 9\delta_\omega$ (black dashed line), whereas the grey dashed line represents the fit to the present data. 
    The ticks at the right margin mark the streamwise wavelengths resolved by the domain length $L_x = 36\delta$.
    }
  \label{fig:lmbx_SCI_domega}
\end{figure}

\citet{Zampiron_secondary_2020} related the enhanced energy at intermediate streamwise wavelengths, which they attributed to a secondary current instability (SCI) of the low-momentum pathways, to the vorticity thickness $\delta_\omega$ of the spanwise shear layers flanking the low-momentum pathways, and found $\lambda_{x,SCI} \approx 9\delta_\omega$ for $\delta \leqslant S \leqslant 2\delta$. Figure~\ref{fig:lmbx_SCI_domega} shows this relation for the present ridge cases, with $\delta_\omega$ defined following \citet{Zampiron_secondary_2020}. 
Here, $\lambda_{x,SCI}$ is the wavelength of the spectral peak of $\kappa_x \Phi_{u''u''}$ at the cell centre $(y_{sc}, z_{sc})$ of the mean secondary motion, and $\delta_\omega$ is evaluated at $y = y_{sc}$. 
For case $Re_\tau = 1000$, $S = 2\delta$, $h = 0.025\delta$, this peak lies in the VLSM range, such that no SCI wavelength can be identified, and this case is excluded from figure~\ref{fig:lmbx_SCI_domega}.

The present data follow a linear relation similar to that of \citet{Zampiron_secondary_2020}, with a larger slope of $\lambda_{x,SCI} \approx 12\,\delta_\omega$. 
Evaluating both quantities at a fixed reference location yields a similar slope (not shown).
This location is the cell centre of case $Re_\tau = 1000$, $S = 2\delta$, $h = 0.1\delta$ ($y_{sc} = 0.453\delta$, $z_{sc} = 0.359\delta$) and lies above the mean secondary motion for small $S$.
The scatter about the fit is, however, considerable and does not reveal a systematic dependence on $S$, $h$ or $Re_\tau$. A substantial part of this scatter results from the discrete streamwise wavelengths resolved in the periodic domain, $\lambda_x = L_x/n$, indicated by the ticks at the right margin of figure~\ref{fig:lmbx_SCI_domega}. 
For a streamwise domain size of $L_x = 36\delta$, adjacent resolved wavelengths differ by 14--20\% for $5\delta \lesssim \lambda_x \leqslant 9\delta$ and by 25\% between $9\delta$ and $12\delta$. 
A small change in the spectral distribution can therefore shift the identified peak to the neighbouring resolved wavelength. 
The present data are thus consistent with a linear relation between $\lambda_{x,SCI}$ and $\delta_\omega$, but do not permit a reliable determination of its slope or of its dependence on the flow and ridge parameters.

%###############################################
\section{Effect of spanwise POD window size}
\label{app-subsec:POD_spanwise_window_size}
%###############################################

\begin{figure}
    \centering
    \includegraphics[]{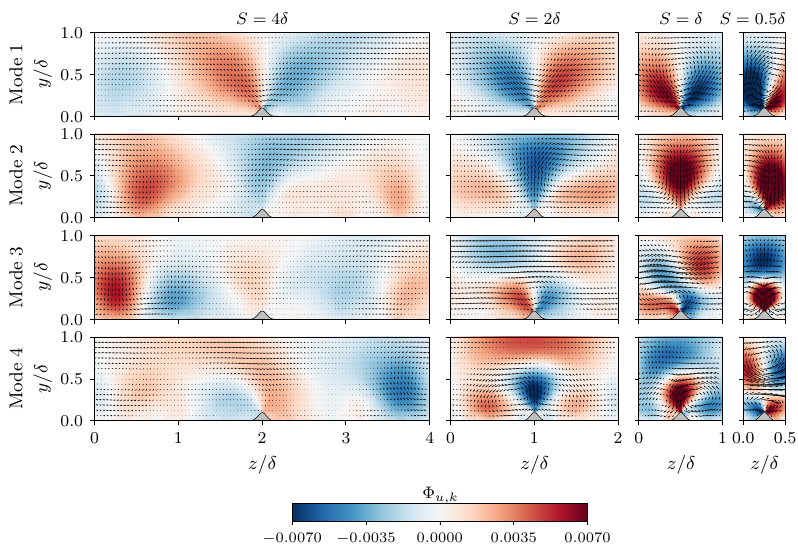}
    \caption{
    First four POD modes for $h=0.1\delta$ at $Re_\tau = 1000$ and different ridge spacing $S$ with POD computation using $L_{z,win} = S$.
    The colour contours show the streamwise velocity component of the POD mode, while the arrows represent the cross-sectional velocity components for the POD mode.    
    }
    \label{fig-app:pod_modes_Lzwin_eq_S}
\end{figure}

The effect of using a spanwise POD window size that scales with the ridge spacing $L_{z,win}=S$ is shown in figure \ref{fig-app:pod_modes_Lzwin_eq_S}.
The first POD mode of all $S$ cases represents the vortical structure centred at the ridge location, which was also found for the POD modes with $L_{z,win}=2\delta$ in figure \ref{fig:pod_modes_S_hg_sweep_Retau1000}.
For $S=4\delta$ it can be seen that this vortical structure has a spanwise extent of roughly $2\delta$, while for the cases $S\le \delta$ this structure is constrained in the spanwise direction by using $L_{z,win}=S$.
The second POD mode represents the large-scale counter-rotating vortices with up- and downward motion above the ridge similar to the one found in figure \ref{fig:pod_modes_S_hg_sweep_Retau1000}.
However, for $S=4\delta$ this structure is less pronounced as this mode also detects a dominant vortical structure in the valley region. 
Similarly, the third and fourth POD modes for $S=4\delta$ capture dominant vortical structures in the valley region, which are not directly related to the secondary motion at the central ridge.
In contrast, the third and fourth POD modes for $S\le2\delta$ detect dominant structures similar to, but smaller than, those found for the first two POD modes, reflecting structures associated with the secondary motion's dynamics.
Overall, this shows that the chosen POD window size of $L_{z,win}=2\delta$ in the POD investigation in \S\,\ref{subsec:pod} is sufficiently large to capture the extent of the dominant structures associated with the instantaneous secondary motions without truncating these structures for small $S$.

%#######################################################################
%\newpage
\bibliographystyle{jfm}
\bibliography{literature.bib}

\end{document}